\documentclass[pre,twocolumn,superscriptaddress,floatfix]{revtex4}
\usepackage{epsfig}
\usepackage{subfigure}
\usepackage{graphicx}
\usepackage{amssymb}
\usepackage{epstopdf}

\begin{document}

\title{Phase diagram of a two-dimensional system with anomalous liquid properties}

\author{Ahmad M. Almudallal}
\affiliation{Department of Physics and Physical Oceanography,
Memorial University of Newfoundland, St. John's, NL, A1B 3X7, Canada}

\author{Sergey V. Buldyrev}
\affiliation{Department of Physics, Yeshiva University,
500W 185th Street, New York, NY 10033, USA}

\author{Ivan Saika-Voivod}
\affiliation{Department of Physics and Physical Oceanography,
Memorial University of Newfoundland, St. John's, NL, A1B 3X7, Canada}

\date{\today}

\begin{abstract}

Using Monte Carlo simulation techniques, we calculate the phase diagram for a 
square shoulder-square well potential in 
two dimensions
that has been previously shown to exhibit liquid anomalies consistent with a metastable liquid-liquid critical point.  
We consider the liquid, gas and five crystal phases, and find that all the melting 
lines are first order, despite a small range of metastability. 
One melting line exhibits a temperature maximum, as well as a pressure maximum that implies inverse melting
over a small range in pressure.

\end{abstract}

\maketitle

\section{Introduction}

Core-softened potentials were first used by Stell, Hemmer and coworkers in a lattice gas system to discuss the isostructural solid-solid phase transition that ends in a second critical point~\cite{hemmer,stell,hoye}. Core-softened potentials were also used to study single-component systems in a liquid state, such as liquid metals~\cite{mon1,selbert,levesque,kincaid, cummings,voronel,mon2,velasco}. They have been also used to study liquid anomalies in 1D~\cite{cho,sadr1,sadr2} and 2D~\cite{jagla1,jagla2,jagla3,sadr3}. Calculations in Ref.~\cite{head1,head2} show that a  core-softened potential can be considered as a realistic first-order approximation for the real interaction between water molecules resulting from averaging over the angular part.

Interest in the study of liquid-liquid (L-L) phase transitions in single component systems grew dramatically after such a transition and accompanying
critical point were proposed for water as an explanation for its anomalous properties~\cite{poole}. 
Various studies have been done to understand the L-L phase transition and associated phenomena. 
Some of these studies focus on the ``two-liquid'' model to explain liquid properties~\cite{rapoport,brazhkin,cuthbertson}. Other studies were based on using anisotropic potentials~\cite{sri,glosli}. Franzese {\it et al.} showed that the liquid-liquid phase transition and accompanying critical point can also arise from an isotropic interaction potential with two characteristic distances (hard-core and soft-core)~\cite{franzese}. 
In this work, the authors reported in 3D molecular dynamics (MD) simulations the existence of two liquid phases, the low-density liquid  phase and the high-density liquid phase, and showed that these two phases can occur in the system with no density anomaly. On the other hand, 2D simulation studies reproduce the density anomaly but no second critical point~\cite{sadr1}.
For a review of unusual behavior of isotropic potentials with two energy scales in 2D, see Ref.~\cite{sergey2}.

\begin{figure}[ht]
\centering\includegraphics[clip=true, trim=0 0 0 0, height=6.0cm, width=8.0cm]{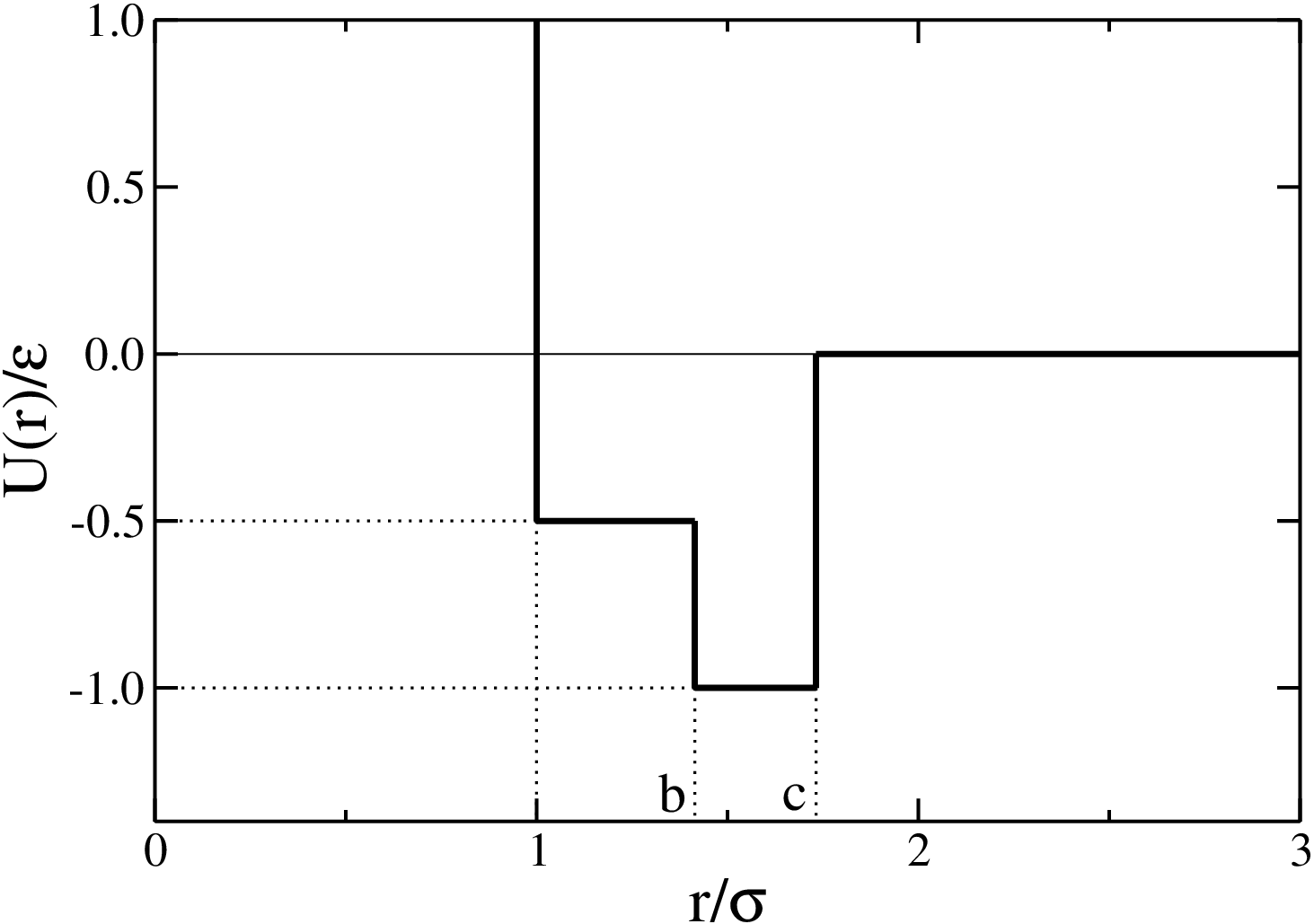}
\caption{The pair potential used in this study is an isotropic step potential  with hard-core diameter $\sigma$.  $b=\sqrt{2}\sigma$ is the soft-core distance, and $c=\sqrt{3}\sigma$ is the attractive distance limit. $r$ is the distance between two particles and $\epsilon$ is the bond energy.}
\label{fig1}
\end{figure}

Scala and coworkers~\cite{sadr3} carried out MD simulations in 2D of 
the square-shoulder square-well (SSSW) potential shown in
Fig.~\ref{fig1} to study liquid anomalies.
Buldyrev {\it et al}~\cite{sergey} continued with the SSSW model in 2D and 3D  in order to study liquid-liquid phase transitions. 
For the 2D system, they produced a phase diagram showing liquid anomalies in relation to approximate crystallization lines for a range in pressure $P$ and temperature $T$ near a potential L-L critical point. 
Their phase diagram shows the gas-liquid coexistence curve and crystallization lines
for a low density triangular 
and higher density square 
crystal. 
It also shows the first critical point and the hypothetical position of the second critical point, which coincides with the 
crossing of the two crystallization lines.  Thus, unavoidable crystallization renders the L-L critical point not directly observable, or obscured.
Their crystallization lines were determined from examining the behavior of the pressure, structure and dynamics along isochores.
They are estimates of the limit of liquid stability, or rather the limit of metastability, with respect to the crystal, rather than thermodynamically 
determined coexistence lines.
As the system is two-dimensional, the nature of the crystallization transition is also under question, in so far that in two dimensions, crystallization can proceed in a continuous way via a hexatic phase rather than through a first-order phase transition.

In the present work, we carry out free energy calculations, based primarily on Monte Carlo (MC) simulation, 
to determine the coexistence conditions between the liquid and crystal phases for a wide range of
$P$ and  $T$, including the smaller range presented in Ref.~\cite{sergey}.  In doing so, we find two low density 
crystal phases not previously reported for the model.  We find that all the transitions are at least weakly first-order.  
The crystallization lines reported in Ref.~\cite{sergey} are below our calculated melting lines.  
Additionally, the square crystal shows a maximum temperature in its melting curve, as well as a maximum in pressure.  
Thus, the present  model is a useful one for studying the rare phenomenon of inverse melting, in which the liquid may freeze to the crystal 
upon heating.

This paper is organized as follows.  In Section II, we discuss all the free energy and computer simulation techniques used in carrying out this work. In section III, we show our results. In Section IV we present a discussion and  we give our conclusions in Section V.

\section{Methods}

\subsection{Model and simulations}

\begin{figure}[htp]
\centering
\subfigure{
\includegraphics[clip=true, trim=0 0 0 0, height=3.6cm, width=3.8cm]{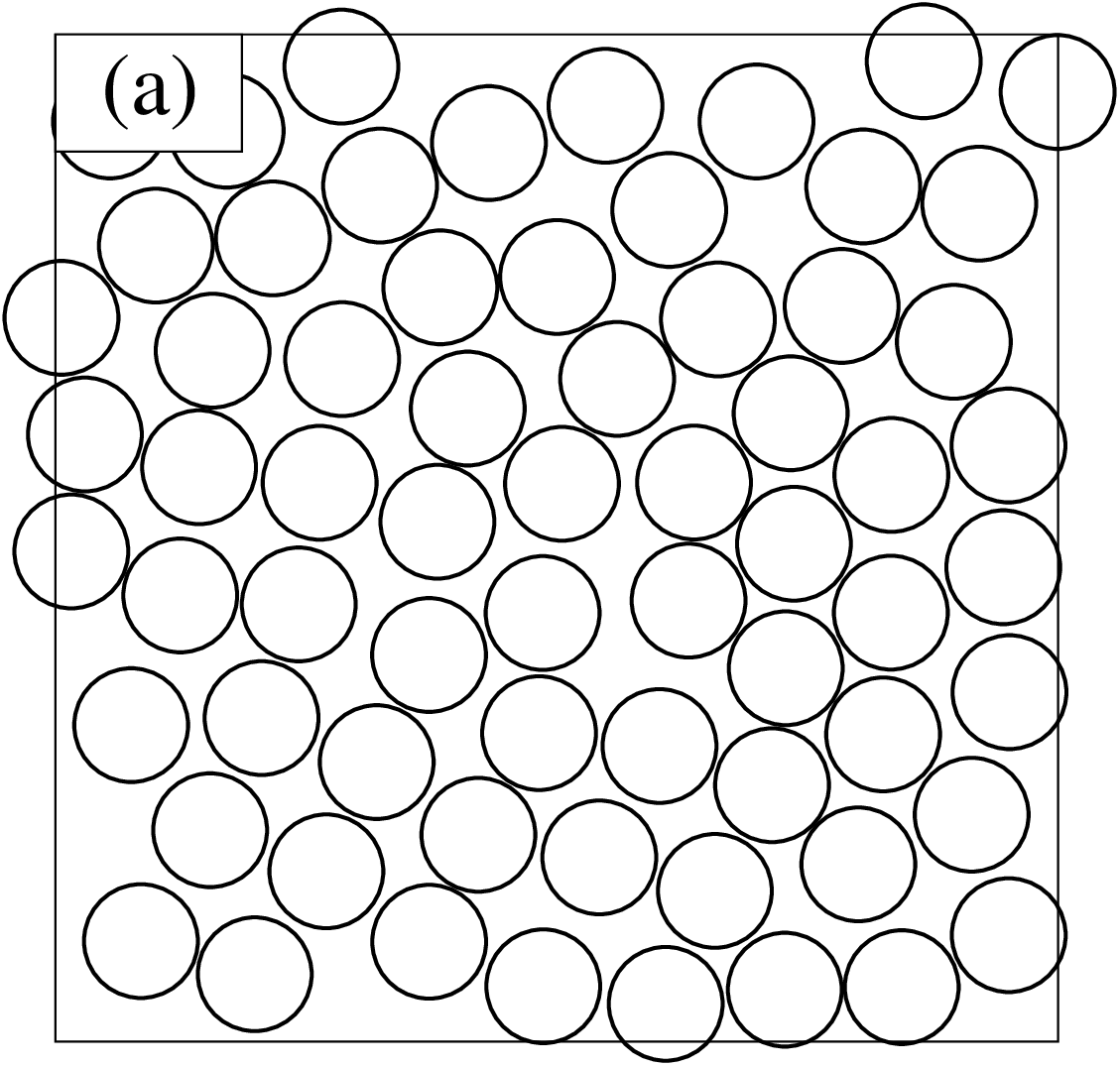}
\label{fig2a}
}
\subfigure{
\includegraphics[clip=true, trim=0 0 0 0, height=3.5cm, width=3.5cm]{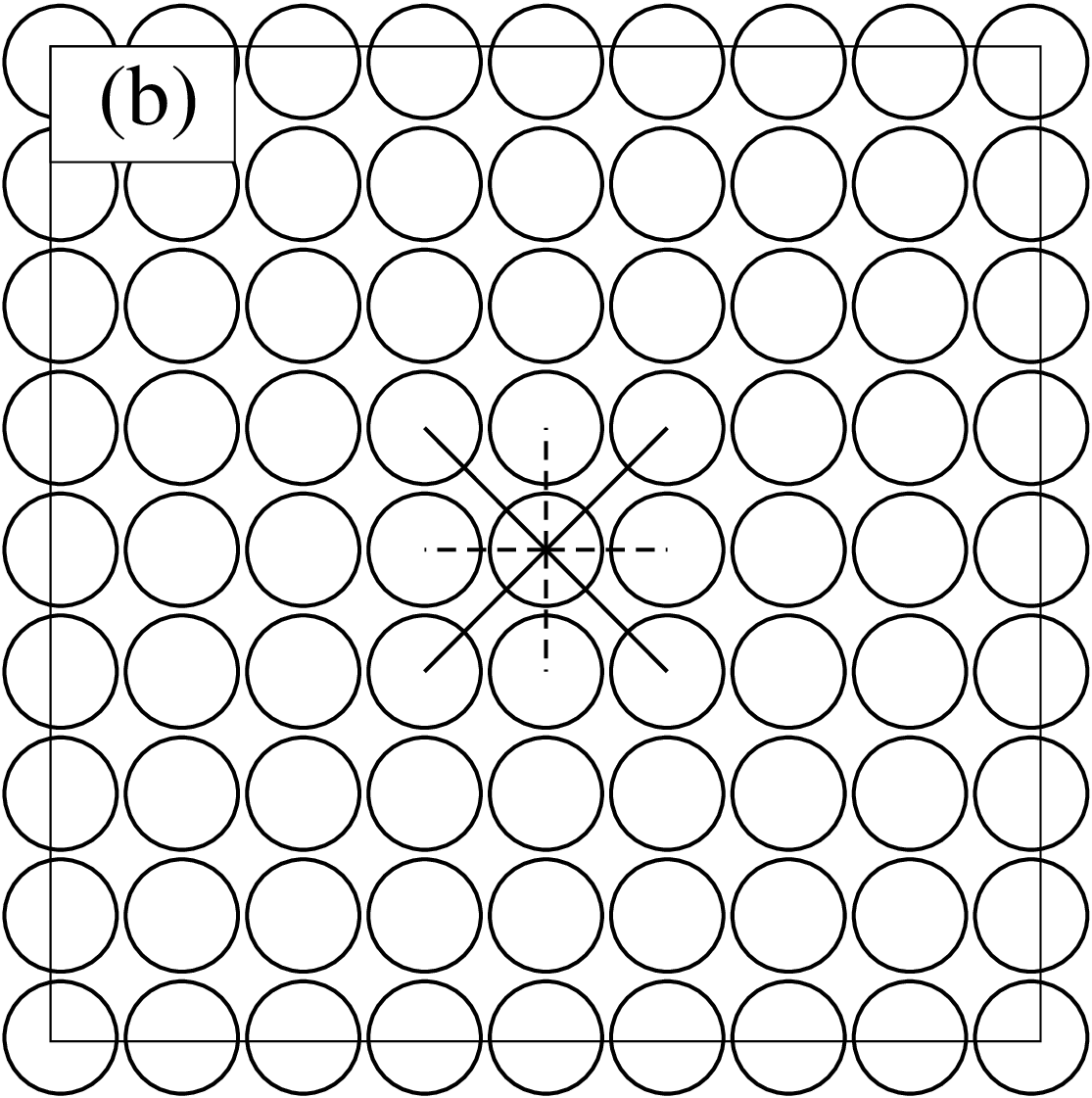}
\label{fig2b}
}
\subfigure{
\includegraphics[clip=true, trim=0 0 0 0, height=3.5cm, width=3.5cm]{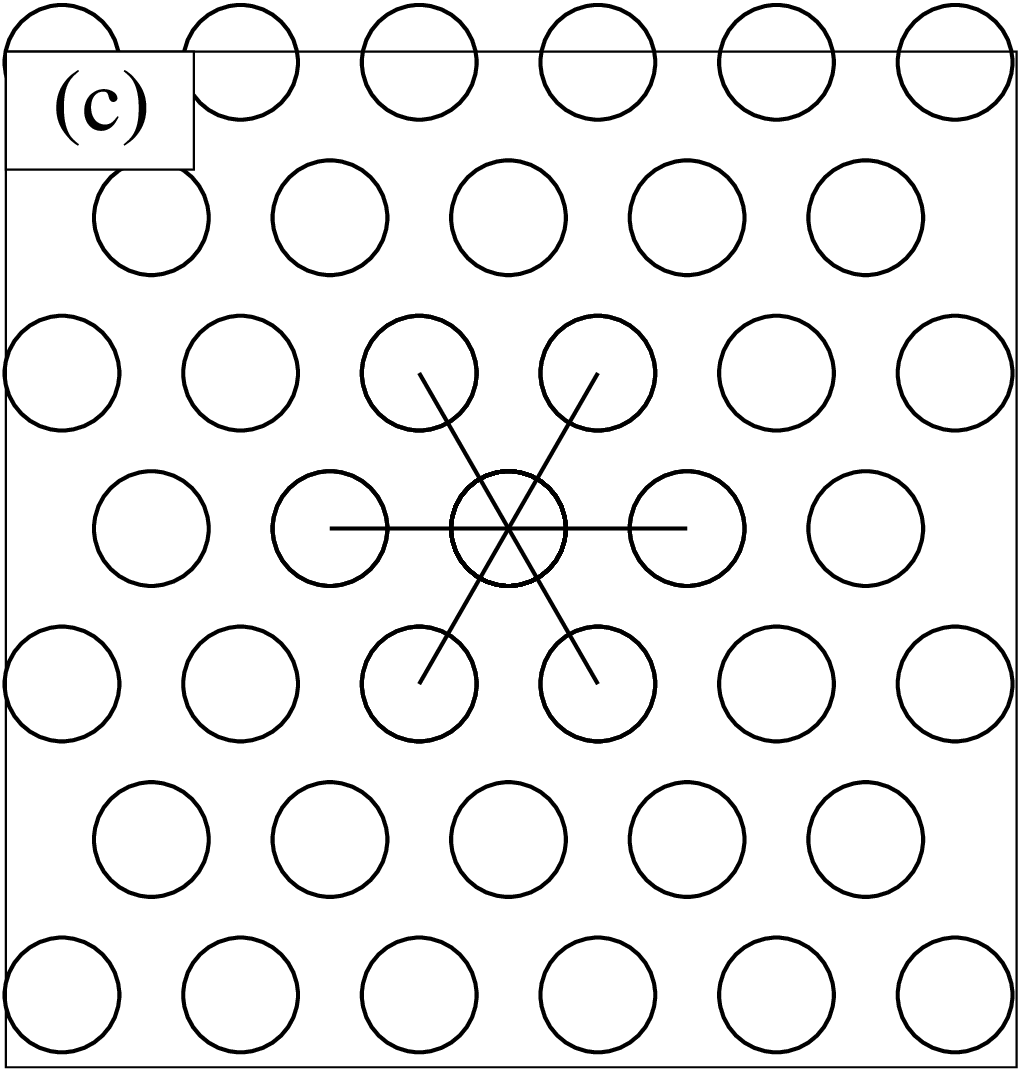}
\label{fig2c}
}
\subfigure{
\includegraphics[clip=true, trim=0 0 0 0, height=3.5cm, width=3.5cm]{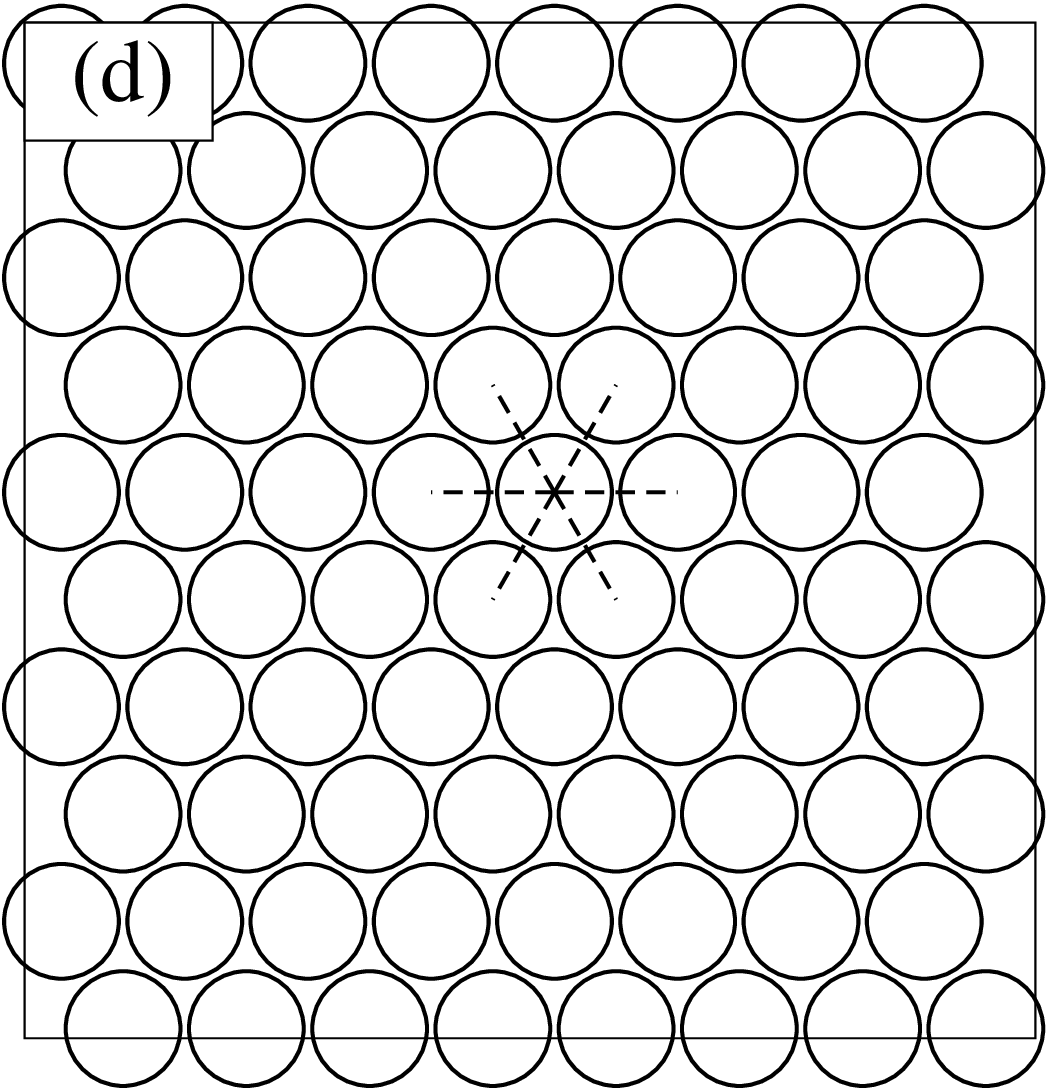}
\label{fig2d}
}
\subfigure{
\includegraphics[clip=true, trim=0 0 0 0, height=3.5cm, width=3.5cm]{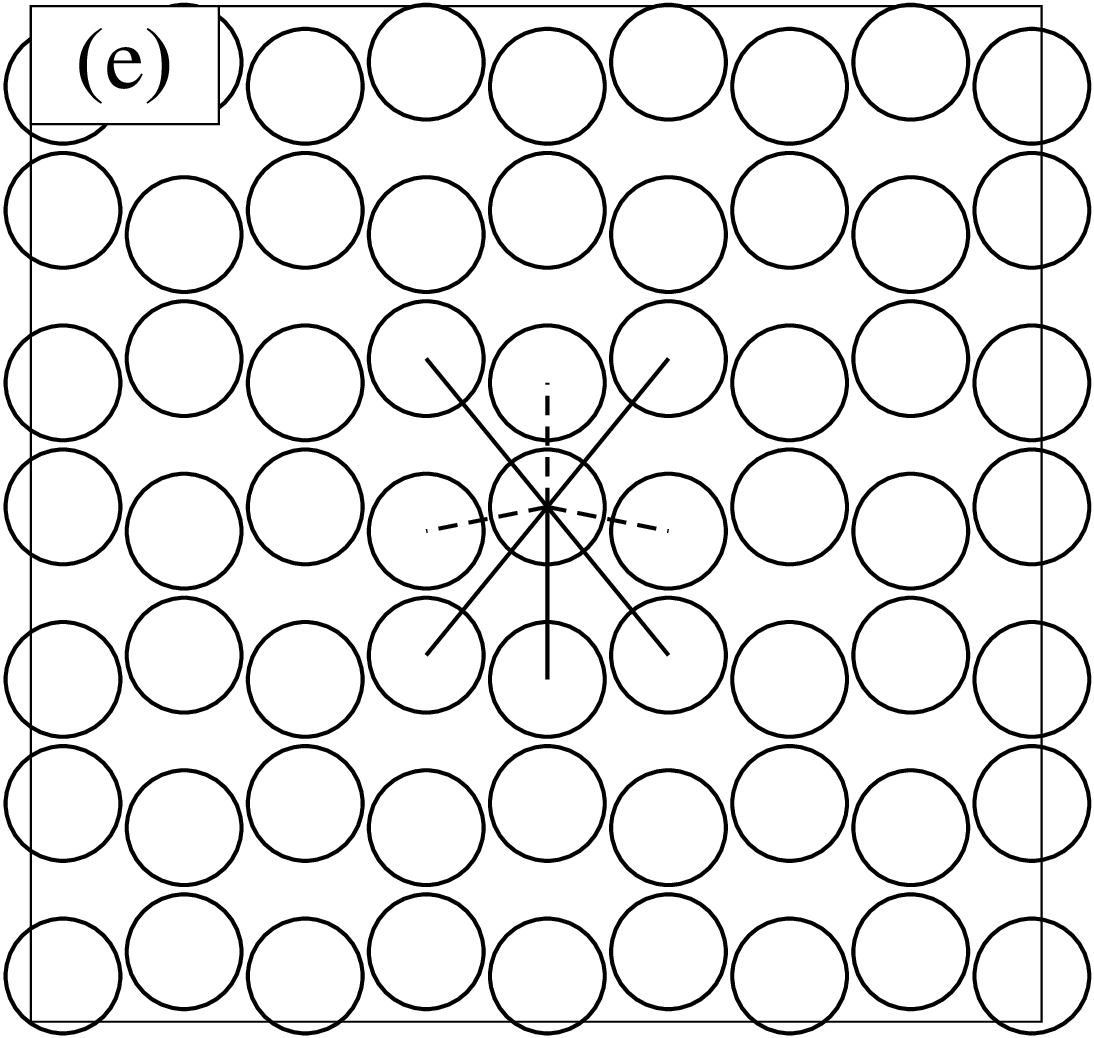}
\label{fig2e}
}
\subfigure{
\includegraphics[clip=true, trim=0 0 0 0, height=3.5cm, width=3.5cm]{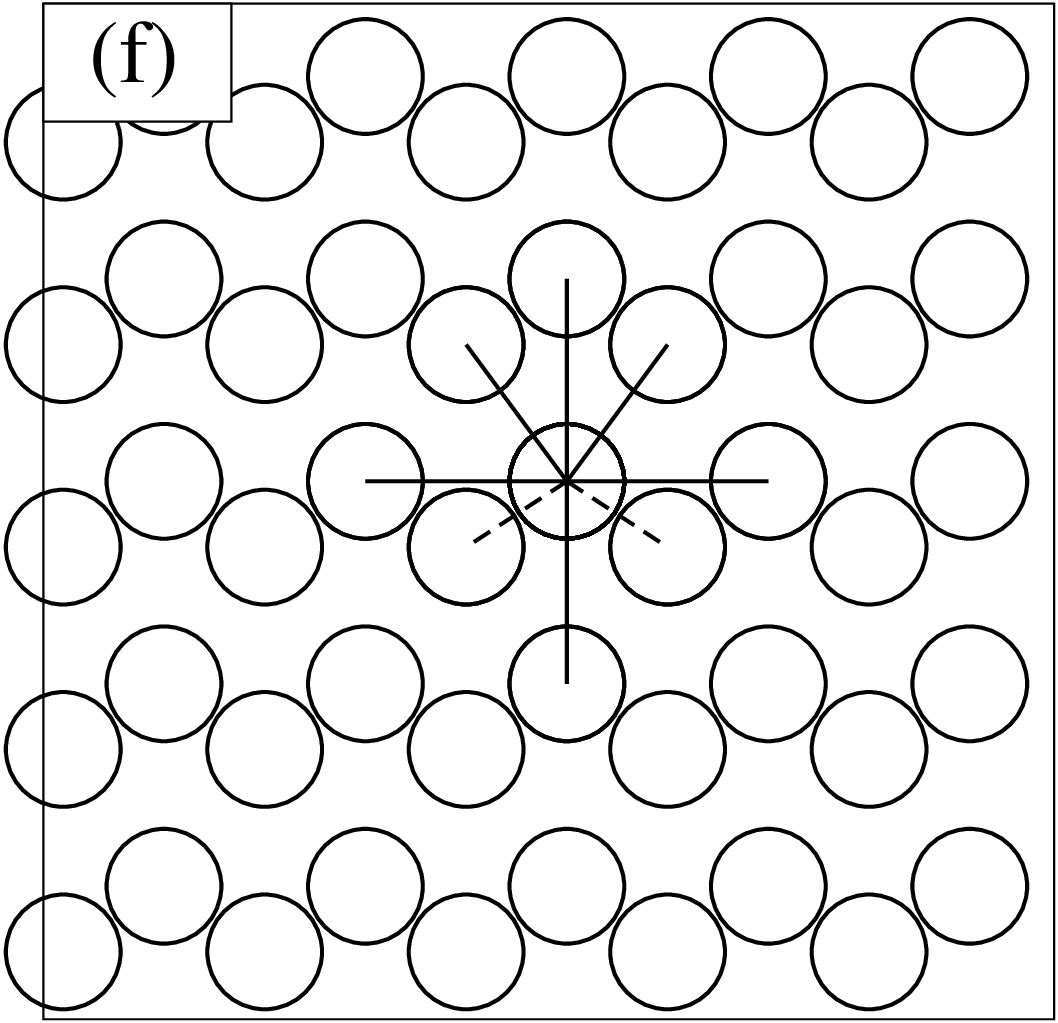}
\label{fig2f}
}
\caption{Illustration of the phases modelled:  \subref{fig2a}
the liquid (L), here shown as a small portion of a simulation  
in which distinct local packing environments are visible, 
\subref{fig2b} the square crystal (S), 
\subref{fig2c} the low-density triangular crystal (LDT), 
\subref{fig2d} the high-density triangular crystal (HDT), 
\subref{fig2e} the A crystal and \subref{fig2f} the Z crystal.
Line segments for the crystal phases indicate a bond with energy $-\epsilon$ and a dashed line segment
one with energy $-\epsilon/2$.}
\label{configurations}
\end{figure}

The model we study is the step pair potential shown in Fig.~\ref{fig1}.  As we are carrying out our studies in two dimensions, the  model
describes disks with a hard-core diameter $\sigma$ and an attractive well extending out to a radial distance of $c=\sqrt{3}\sigma$. The attractive well itself contains
a shoulder, with a pair interaction energy of $-\epsilon/2$ for $\sigma < r < b$ and energy of $-\epsilon$ for $b < r < c$.  The parameter $b$ was originally chosen to be
$\sqrt{2}\sigma$ so that there would exist a low density triangular (LDT) phase  and a higher density square (S) phase with the same 
potential energy per particle of $-3\epsilon$, i.e. two energetically 
degenerate phases of well separated densities~\cite{sergey}.  
The idea behind this was to allow for distinct liquid states, one based on square packing and the other on 
the more open triangular packing, in analogy to what is thought to be the case for water.  At high pressure the system ultimately must form the 
close-packed triangular phase (HDT), with potential energy per particle of $-1.5\epsilon$.  We find two additional crystals,
phases A and Z, with per particle energies $-3.25\epsilon$ and $-3.5\epsilon$, respectively.  The various crystal phases are depicted in 
Fig.~\ref{configurations}.  Our goal is to calculate coexistence lines between the five crystal phases, the liquid (L) and the gas (G).

The liquid-state properties of the model were extensively studied in Ref.~\cite{sergey} using discrete MD simulation.
The S and LDT crystallization lines were determined in that work from pressure isochores and from direct observation of crystal-like
structural and dynamical behavior.  Here, we calculate the crystal coexistence lines using free energy techniques that employ for the most
part MC simulations performed at constant particle number $N$, $P$ and  $T$, i.e., in the NPT ensemble~\cite{frenkel1}.  
Depending on the phase, the pressure is kept constant by
changing the volume isotropically (for L, S,  HDT and LDT), 
by allowing rectangular dimensions of the simulation cell to change length independently while
maintaining a right angle (for A and Z), or by allowing the angle to change as well (as a check for all phases).
The system sizes and box shapes are as follows: 
(L) $N=1020$ and $986$, square box; 
(S) $N=1024$ square box, and $N=992$ with rectangular box $L_y = 32 L_x/31$; 
(HDT and LDT) $N=986$, $L_y=17 \sqrt{3}L_x/29 $; 
(A) $N=952$, $L_y=28(\sin12^{\circ}+1) L_x/(34 \cos12^{\circ})$ initially; 
(Z) $N=968$, square box initially.
The different box shapes (and hence number of particles) are used as consistency checks, and indeed 
we do not detect any difference in the results based on  the particular choice used.

\subsection{Solid-liquid and solid-solid coexistence}
\label{sol-liq-sec}

First-order transition lines can be determined using a method developed by Kofke to trace coexistence curves~\cite{kofke1,kofke2}. Kofke refers to his method as Gibbs-Duhem integration, and it is based on the Clapeyron equation which describes the temperature dependence of the pressure at which two phases coexist,
\begin{equation}
 \frac{d P}{d T} = \frac{\Delta s}{\Delta v} = \frac{\Delta h}{T \Delta v},
\label{gibbsduhem}
\end{equation}
where  $\Delta s$ is the molar entropy difference, $\Delta h$ is the molar enthalpy difference and $\Delta v$ is the molar volume difference between the two coexisting phases. Tracing the coexistence curve requires that one point on the coexistence curve be known and then the rest of the curve can be found by integration of Eq.~\ref{gibbsduhem}, in particular using the enthalpy since it is much easier to calculate than the entropy. 
We carry out the integration using a second-order predictor-corrector method~\cite{lamm,predcorr}.

To obtain the first coexistence point between the liquid and the S crystal, we first determine the respective equations of state along an
isotherm by carrying out several $NPT$ simulations. We choose $k_B T/\epsilon=0.55$ so that we are above the L-G critical temperature, where $k_B$ is the Boltzmann constant. Once the equations of state are known, we calculate the chemical potential $\mu$ for each phase as a function of number density $\rho$ by integrating 
the pressure via~\cite{frenkel1,vega1},
\begin{equation}
\label{muintegral}
\beta \mu(\rho) = \beta f(\rho^*) + \beta \int_{\rho^*}^{\rho} \frac{P(\rho^\prime)}{{\rho^\prime}^2} d\rho^\prime + \frac{\beta P}{\rho},
\end{equation}
where $\beta=(k_B T)^{-1}$ and $f$ is the Helmholtz free energy per particle calculated at a reference number density $\rho^*$.

To carry out the integration, we fit the liquid isotherm to Eq.~\ref{liq-eq-state} and the solid isotherm to Eq.~\ref{sol-eq-state}~\cite{pagan,massimo},

\begin{eqnarray}
 \beta P &=& \frac{\rho}{1- a_l\rho} + b_l \left( \frac{\rho}{1- a_l\rho} \right)^2 + c_l \left( \frac{\rho}{1- a_l\rho} \right)^3  \label{liq-eq-state}, \\
 \beta P &=& a_s \rho^2 + b_s \rho + c_s,  \label{sol-eq-state}
\end{eqnarray}
where $a_{l,s}$, $b_{l,s}$, and $c_{l,s}$ are the fit parameters. Integration of Eq.~\ref{liq-eq-state} from zero to a density of interest yields the chemical potential of liquid, as given in Eq.~\ref{chem-liq}. Similarly, integration of Eq.~\ref{sol-eq-state} from a reference density to the density of interest yields the chemical potential of solid, as given in Eq.~\ref{chem-sol}~\cite{pagan,massimo},
\begin{eqnarray}
 \label{chem-liq}
 \beta \mu_l(\rho) &=& \ln\left( \frac{\rho \Lambda^2}{1 - a_l\rho} \right) + \frac{b_l/a_l - c_l/a_l^2 + 1}{1 - a_l\rho} \nonumber \\
                   &+& \frac{c_l/2a_l^2 + b_l\rho}{(1 - a_l\rho)^2}    + \frac{c_l\rho^2}{(1 - a_l\rho)^3}  \nonumber\\
                   &-& (b_l/a_l - c_l/2a_l^2 + 1), \\
 \nonumber \\
 \beta \mu_s(\rho) &=& 2a_s\rho + b_s[\ln(\rho) + 1] \nonumber \\
                   &-& [a_s\rho^{*} + b_s\ln(\rho^{*}) - c_s/\rho^{*}] \nonumber \\
                     &+& \beta f^{\rm ex}(\rho^{*}) + \ln(\Lambda^2 \rho^{*}) - 1,   \label{chem-sol}
\end{eqnarray}
where $\Lambda = h/\sqrt{(2 \pi m k_B T)}$ is the de Broglie thermal wavelength, where it is assumed to equal unity  since it plays no rule in locating the coexistence pressure (along an isotherm). $f^{\rm ex}(\rho^{*})$ is the excess Helmholtz free energy per particle calculated at  $\rho^*$. 

\begin{figure}[ht]
\centering\includegraphics[clip=true, trim=0 0 0 0, height=6.0cm, width=8.0cm]{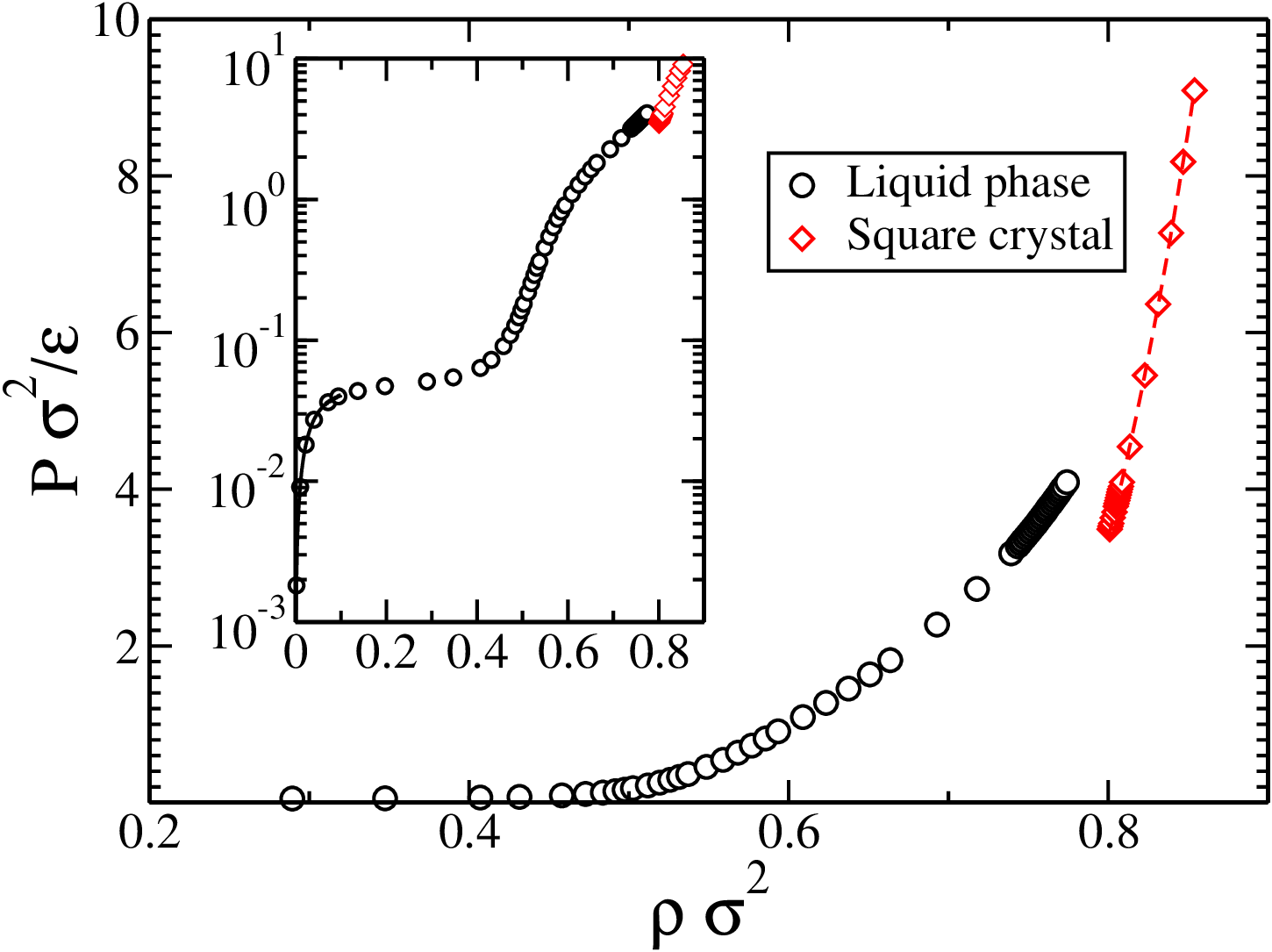}
\caption{Equations of state of the liquid (circles) and S crystal (diamonds) at $k_B T/\epsilon=0.55$.  The curves show fits according
to Eqs.~\ref{liq-eq-state} and \ref{sol-eq-state}, with $a_l=-0.457818$, $b_l=-9.18372$, and $c_l=33.4007$ for the liquid (inset) and
$a_s=479.035$, $b_s=-686.583$, and $c_s=246.067$ for the crystal.}
\label{fig3}
\end{figure}

For the liquid, Eq.~\ref{liq-eq-state} provides a good fit only up to $\rho\approx 0.1$, and so from $\rho=0$ to $\rho^*_l=0.09418$ we use Eq.~ \ref{chem-liq},
and then integrate Eq.~\ref{muintegral} numerically, using different interpolation orders to estimate uncertainty.  The equations of state for the liquid and the S crystal are shown in Fig.~\ref{fig3}.

\begin{figure}[ht]
\centering\includegraphics[clip=true, trim=0 0 0 0, height=6.0cm, width=8.0cm]{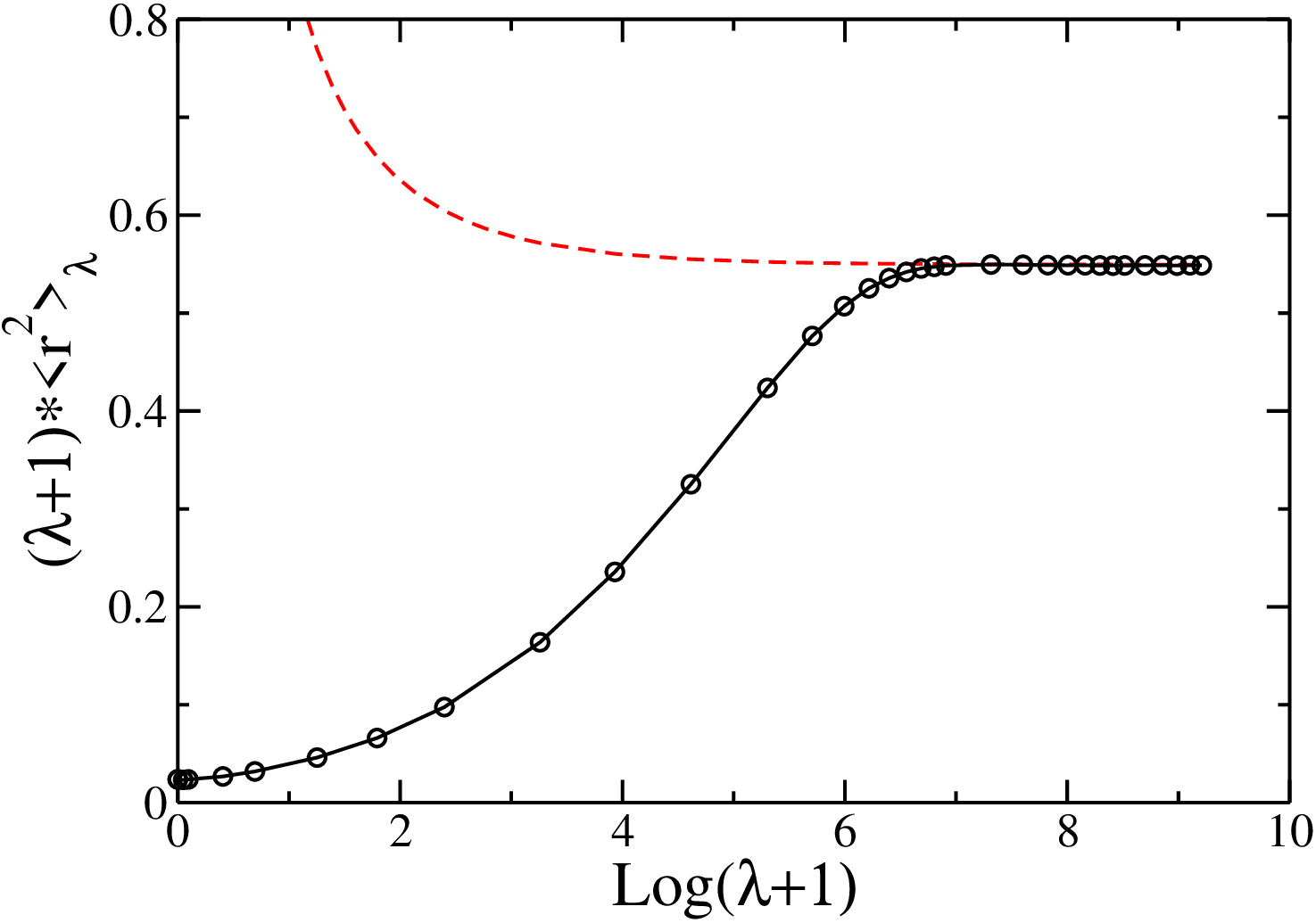}
\caption{The mean-squared displacement transformed by Eq.~\ref{transformed_integration} as a function of coupling parameter $\lambda$ calculated by computer simulation (solid curve is a guide to the eye).  Dashed line is the theoretical value given by Eq.~\ref{mean_sqre_displ_theory}.}
\label{fig4}
\end{figure}

We calculate the crystal reference Helmholtz free energy using the Frenkel-Ladd  method~\cite{frenkel2}. In this method,
a harmonic potential is added to the original system to define a new system potential energy,
\begin{equation}
U_{\lambda}=U(\vec{r}^N) + \lambda \sum_{i=1}^N (\vec{r}_i - \vec{r}_{0,i})^2,
\end{equation}
where $\vec{r}_i$ is the position of particle $i$ and $\vec{r}_{0,i}$ is its ideal lattice position, and 
$U(\vec{r}^N)$ is the unaltered system potential energy.
$U_{\lambda}$ is such that at coupling parameter $\lambda=0$ the original 
model is recovered and for sufficiently large $\lambda$, the system behaves as 
an ideal Einstein crystal.
A thermodynamic integration at a particular $T$ and $\rho$
is carried along $\lambda$ to determine the Helmholtz free energy difference between the Einstein crystal and the original model.
The excess free energy per particle for the model is then expressed as~\cite{frenkel1},
\begin{eqnarray}
 \beta f^{\rm ex} &=&  \beta f_{\rm Ein} + \frac{\beta \Delta F^{CM}}{N} + \frac{\ln(\rho^*)}{N} - \frac{d}{2N}\ln(N) \nonumber \\
                  &-& \frac{d}{2N} \ln\left(\frac{\beta \lambda_{\rm max}m}{2\pi}\right) - \beta f^{\rm id},
\label{external_free_energy}
\end{eqnarray}
where $d=2$ is the dimensionality of the system, $m=1$ is the mass of the particle.
The first term in Eq.~\ref{external_free_energy} represents the free energy of the ideal (non-interacting) Einstein crystal, which is equal to,
\begin{equation}
 \beta f_{\rm Ein} = \frac{\beta U(\vec{r}_0^N)}{N} - \frac{d}{2\beta}\ln\left(\frac{\pi}{\beta\lambda_{\rm max}}\right),
\end{equation}
where $U(\vec{r}_0^N)$ is the potential energy of the crystal when all the atoms are at their ideal lattice positions, and $\lambda_{\rm max}$ is chosen such that, for $\lambda$ larger than $\lambda_{\rm max}$, the mean-squared displacement 
$\langle \delta r^2 \rangle_{\lambda} \equiv  \langle (\vec{r}_i - \vec{r}_{0,i})^2 \rangle_{\lambda}$,
where $\langle \dots � \rangle_{\lambda}$ indicates an ensemble average, 
for a system with  fixed center of mass follows the following analytical expression,
\begin{equation}
 \langle \delta r^2\rangle_{{\rm Eins},\lambda} =  \frac{N-1}{N}\frac{1}{\beta \lambda}.
\label{mean_sqre_displ_theory}
\end{equation}
The second term in Eq.~\ref{external_free_energy} represents the free energy difference between the solid and the Einstein crystal,
 and can be calculated by integrating the mean-squared displacement obtained from simulations 
 carried out with a fixed center of mass as follows~\cite{frenkel1, vega2},
\begin{equation}
 \frac{\Delta F^{\rm CM}}{N} = \int_{0}^{\lambda_{\rm max}} \langle \delta r^2\rangle_{\lambda} d\lambda.
\label{mean_sqre_integral}
\end{equation}
This integration can be understood as gradually switching on the coupling parameter to transform the solid into an Einstein crystal. 
For better accuracy, this integral can be transformed to~\cite{frenkel1},
\begin{equation}
 \frac{\Delta F^{\rm CM}}{N} = \int_{\ln(c)}^{\ln(\lambda_{\rm max}+c)} d[\ln(\lambda+c)](\lambda+c)\langle r^2\rangle_{\lambda},
\label{transformed_integration}
\end{equation}
where $c$ is a constant chosen to be 1 in this work.  The integrand is shown in Fig.~\ref{fig4}, along with 
the curve for the ideal solid. We choose $\ln{(\lambda_{\rm max}+1)}=6.909$, checking that using higher values yields no 
appreciable change in the final result. The integration is carried out using interpolations of different order in order to estimate uncertainty.

The third, fourth and fifth terms in Eq.~\ref{external_free_energy} correspond to the difference between the constrained (fixed center of mass) and unconstrained (non-fixed center of mass) solids. The last term in Eq.~\ref{external_free_energy} is the free energy of the ideal gas
per particle, which is given by,
\begin{equation}
\beta f_{\rm id} = \ln(\rho) - 1 + \frac{\ln(2\pi N)}{2N}.
\end{equation}

Once the chemical potentials of the two phases are known, the coexistence point can be obtained from the intersection of the two chemical potential curves~\cite{massimo,pagan}. 

\begin{figure}[ht]
\subfigure{
\centering\includegraphics[clip=true, trim=0 0 0 0, height=6.0cm, width=8.0cm]{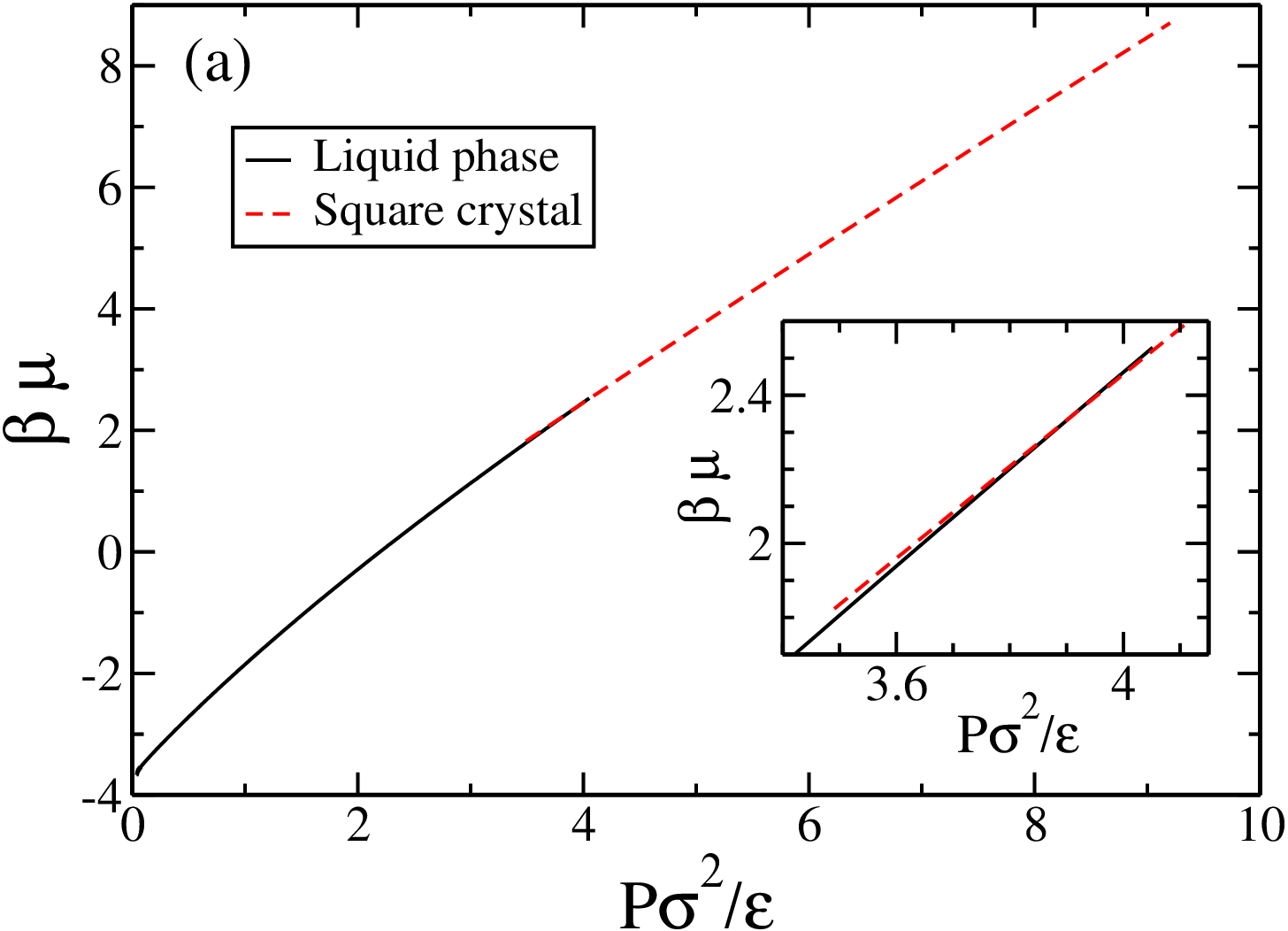}
}
\subfigure{
\centering\includegraphics[clip=true, trim=0 0 0 0, height=6.0cm, width=8.25cm]{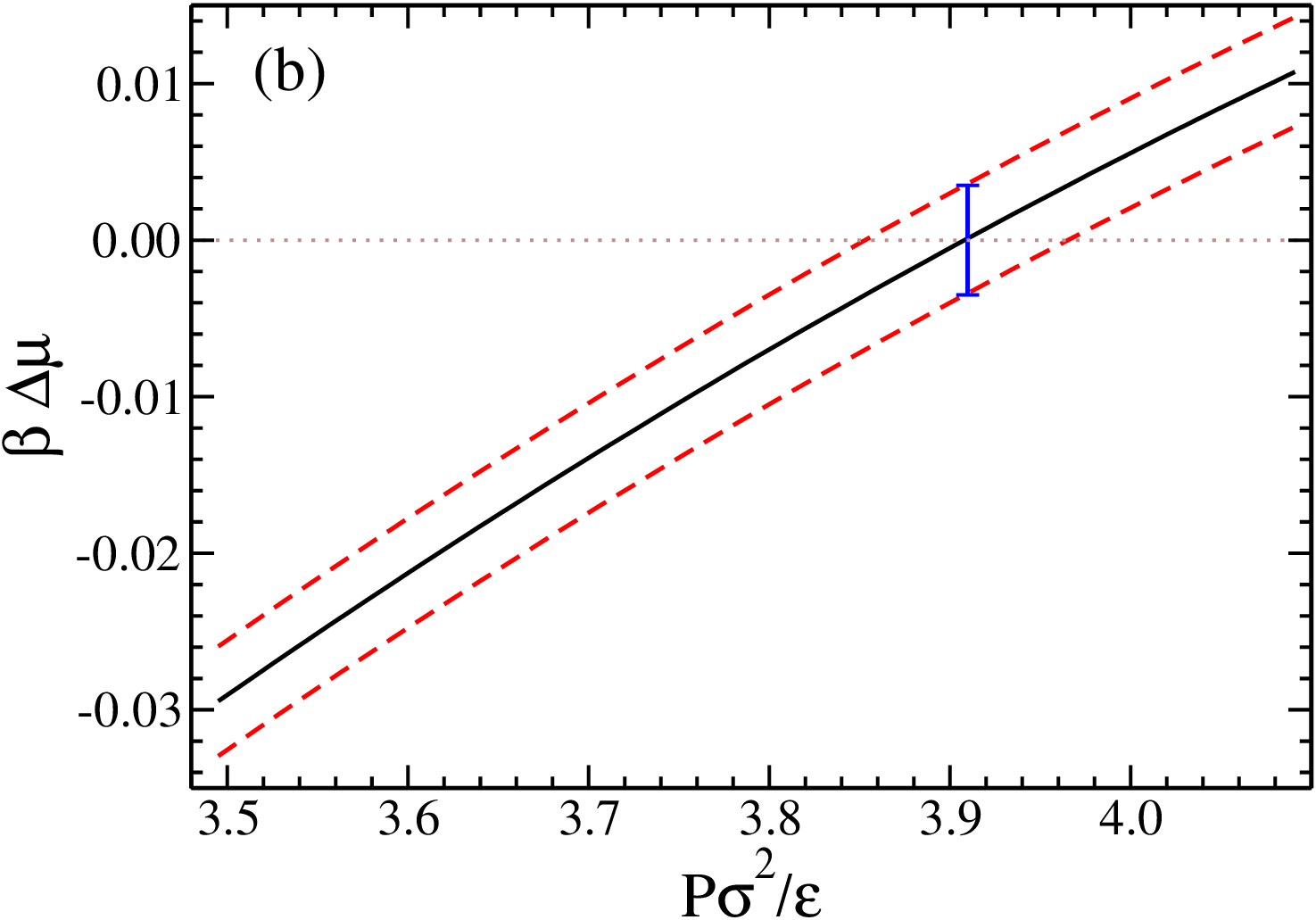}
}
\caption{
Determination of a coexistence $P$ between the L and S phases at $k_B T/\epsilon = 0.55$.
Panel (a) shows the chemical potential isotherms for the liquid (solid curve) and the square crystal (dashed curve).
Inset shows a close-up of the crossing.
In panel (b) we show the difference in chemical potential $\Delta \mu$
between the two phases over the entire range of $P$ for which the equations of state overlap, 
with dashed lines indicating upper and lower uncertainty estimates.}
\label{fig5}
\end{figure}

$\mu_l(\rho)$ and $\mu_s(\rho)$ are used together with the equations of state to plot the chemical potentials of the two phases as functions of pressure,
as we do in Fig.~\ref{fig5}.  It is immediately apparent that $\mu(P)$ has nearly the same slope for both phases, and hence the location of the 
crossing is sensitive to errors in the various calculated quantities used to determine the curves.  
We note that the equations of state are determined only to the point where the metastable phase does not easily transform to the other phase.
It is somewhat surprising that at the  $P$ for  which either phase becomes unstable, $P \sigma^2/\epsilon \sim 3.49$ for S and $P \sigma^2/\epsilon \sim 4.09$
for L, the difference in chemical potential is very small, on the order of  $|\beta \Delta \mu|  \sim 0.01$.

\begin{figure}[ht]
\centering\includegraphics[clip=true, trim=0 0 0 0, height=8.0cm, width=8.0cm]{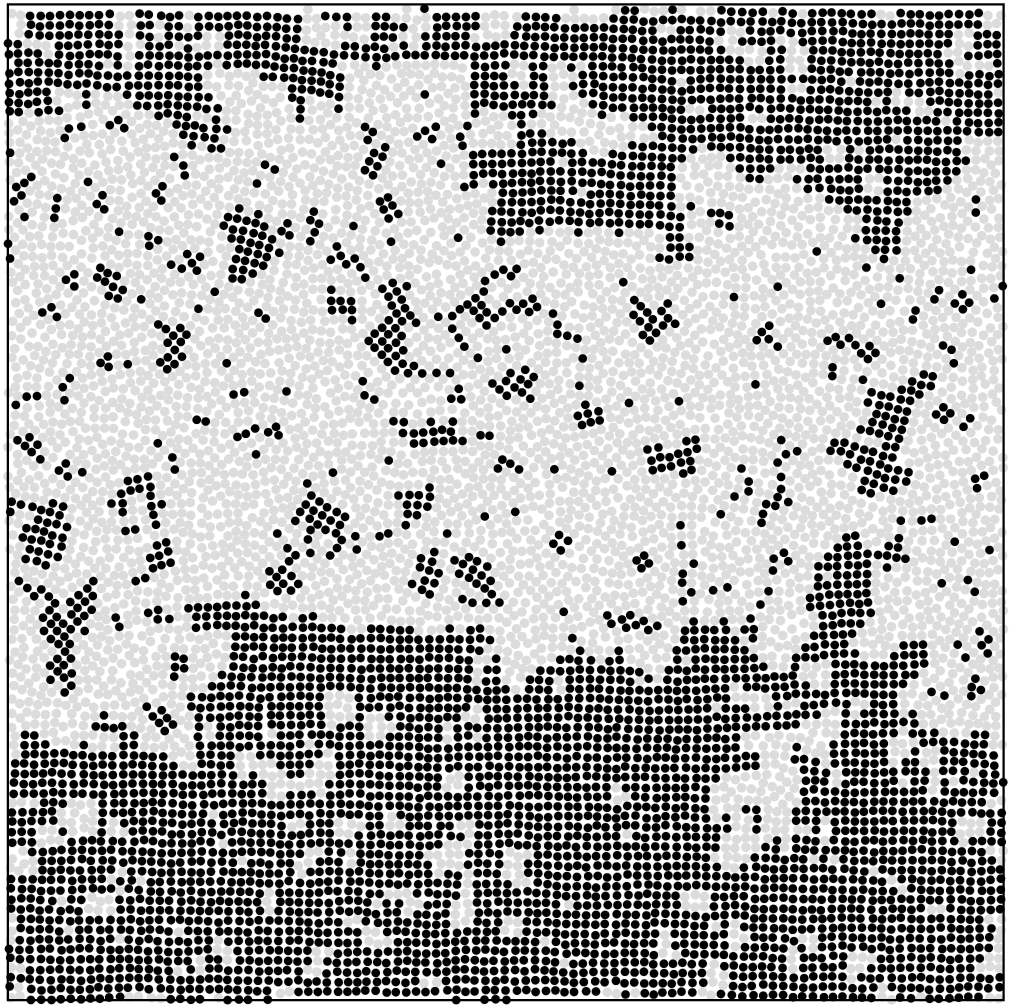}
\caption{Snapshot configuration obtained from an NVT simulation for 10,000 particles at $k_B T/\epsilon=0.55$ and $\rho=0.786567$. Black symbols 
represent particles belonging to the S phase, while grey symbols represent the L phase.}
\label{fig6}
\end{figure}

As a check on the L-S coexistence conditions at $k_B T/\epsilon = 0.55$, we perform an $NVT$ (canonical ensemble) 
simulation with 10,000 particles initially placed
on a square lattice with $\rho \sigma^2=0.786567$, the $\rho$ at which the system is expected to phase separate into L and S with equal numbers
of particles in each phase, based on liquid and S coexistence densities of 
$\rho_l=0.7677$ and $\rho_x=0.8064$, respectively.
Fig.~\ref{fig6} shows a snapshot after running for $2\times10^7$ MC steps per particle, with dark symbols identifying particles belonging to the S phase~\cite{identifysolid,frenkel3,ahmad,ivan}. 
Averaging over the last $5\times10^6$ MC steps per particle, the fraction of particles belonging to the S phase is $0.51$.

The above procedure is repeated (at lower $T$) for the other crystal phases to determine crystal-crystal coexistence lines.
For two crystals, the slopes of  $\mu(P)$ are generally quite different, which makes it easier to pinpoint the coexistence $P$.
Similarly, at $T$
less than the L-G critical temperature, the procedure is repeated to find crystal sublimation lines after determining the equation of state 
for the gas.

\begin{figure}[!ht]
\subfigure{
\centering\includegraphics[clip=true, trim=0 0 0 0, height=6.0cm, width=8.0cm]{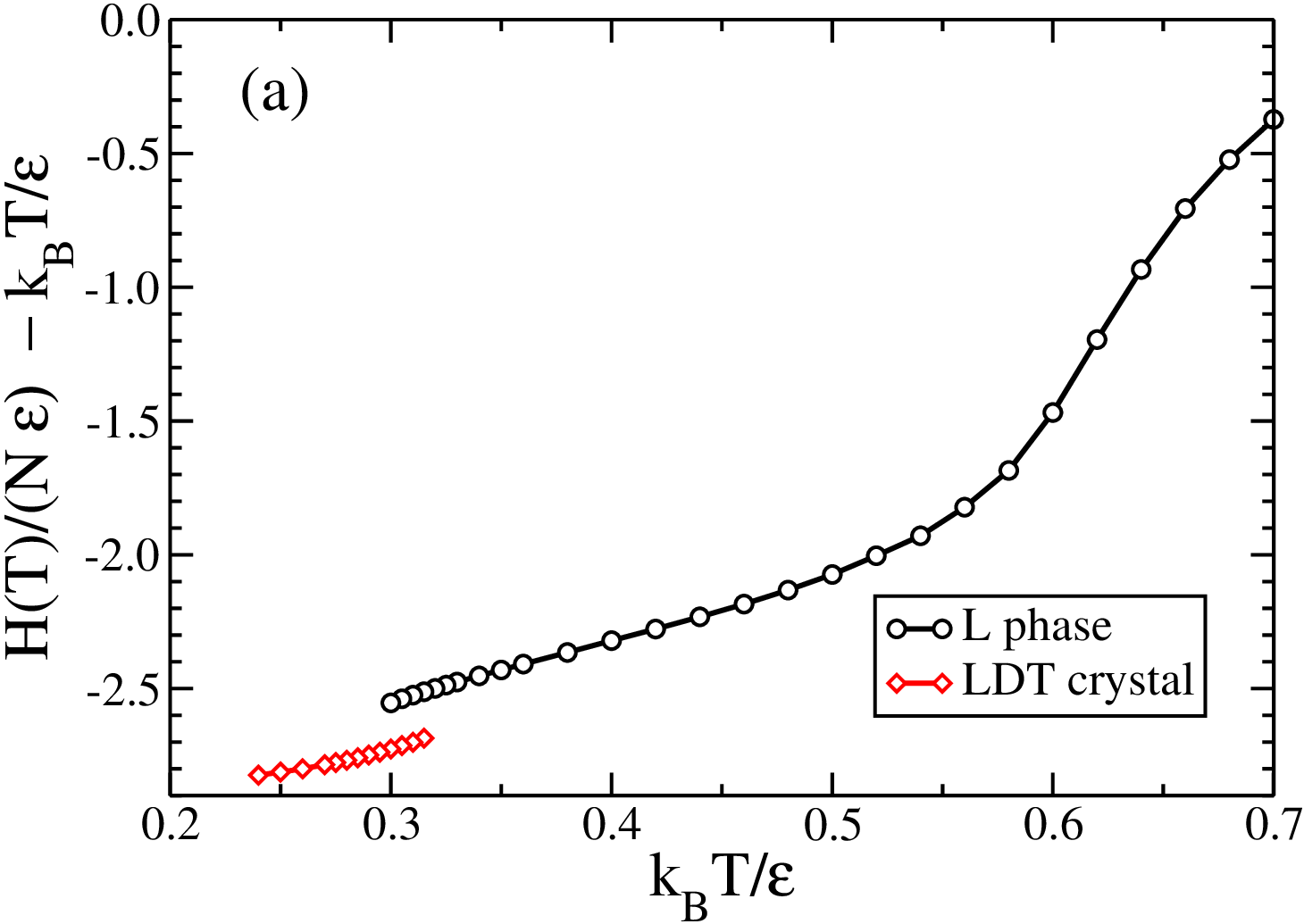}
}
\subfigure{
\centering\includegraphics[clip=true, trim=0 0 0 0, height=6.0cm, width=8.0cm]{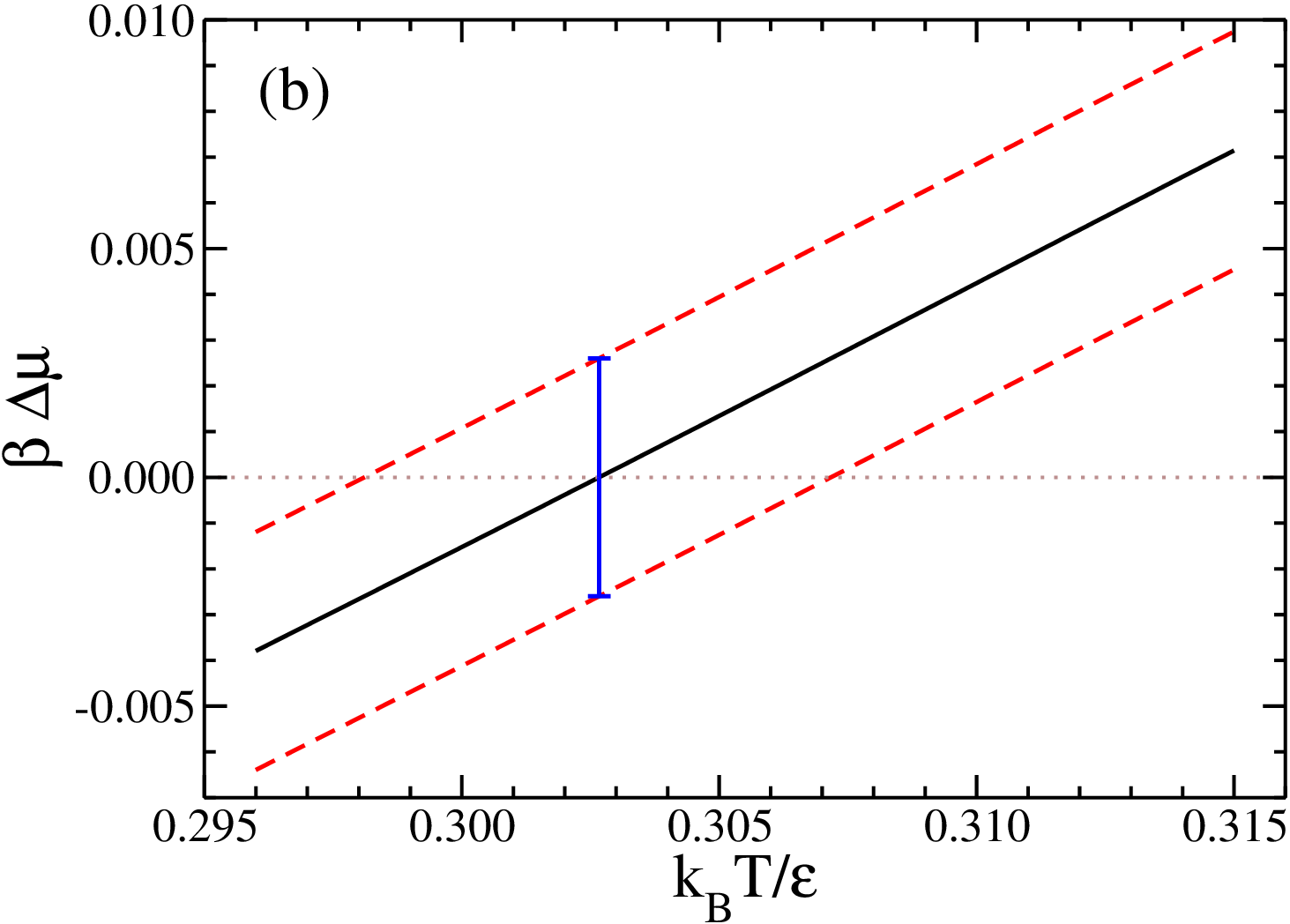}
}
\subfigure{
\centering\includegraphics[clip=true, trim=0 0 0 0, height=6.0cm, width=8.0cm]{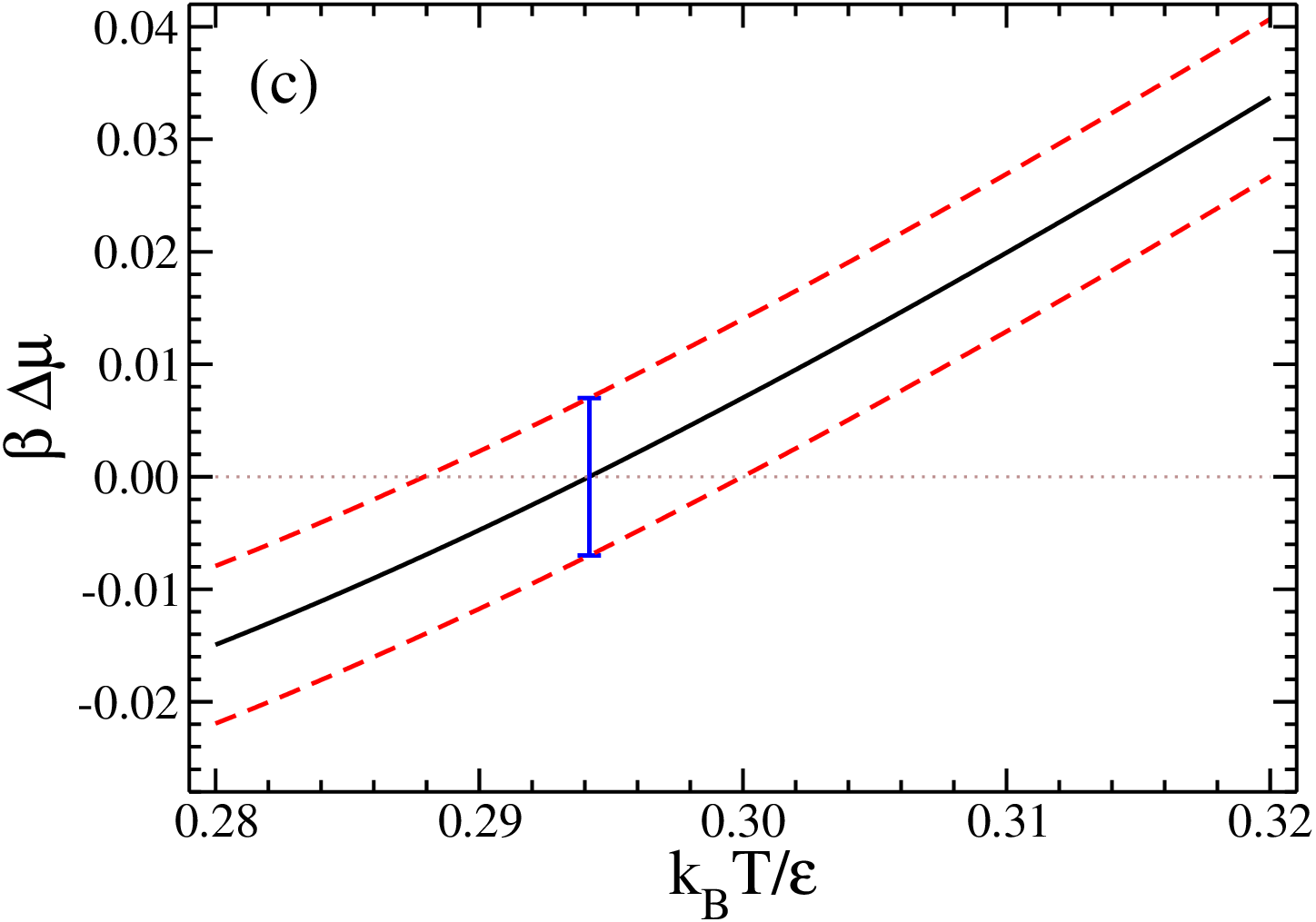}
}
\caption{(a) Enthalpy per particle for the liquid (circles) and LDT crystal (diamonds) along $P\sigma^2/\epsilon=0.05$.  
Here we have subtracted the ideal gas contribution to the  energy.  (b) The corresponding chemical
potential difference between the L and LDT  phases for the entire range in $T$ of metastability.
(c) The chemical potential difference between the L and S  phases at $P\sigma^2/\epsilon=0.15$.}
\label{fig7}
\end{figure}

For the L-LDT melting line, we must additionally perform an integration of the enthalpy $H$ to lower $T$ at a $P$ above the critical pressure 
in order to avoid the L-G critical point. Specifically, we first integrate the liquid equation of state at $k_B T/\epsilon=0.70$ using Eq.~\ref{muintegral} to 
$P \sigma^2/\epsilon = 0.05$, and then calculate $\mu(T)$ via~\cite{vega2},
\begin{equation}
\frac{\mu(T_{2},P)}{k_B T_{2}} = \frac{\mu(T_{1},P)}{k_B T_{1}} - \int_{T_{1}}^{T_{2}} \frac{H(T)}{N k_B T^2} dT,
\label{free_energy}
\end{equation}
noting that here, the $T$ dependence of $\Lambda$ must be taken into account.  Equivalently, this amounts to using the potential energy 
instead of the thermal energy in calculating $H$.  For the LDT crystal, the reference free energy is calculated at $P \sigma^2/\epsilon = 0.05$
after determining the density at that pressure to be $\rho \sigma^2 = 0.4780\pm0.0015$. In Fig.~\ref{fig7} we show $H(T)$ for L and S as well as
the resulting difference in $\mu$ between the phases.  We repeat the calculation using the 
the liquid equation of state at $k_B T/\epsilon=0.55$ as a check.  
Using the same procedure at $k_B T/\epsilon=0.70$, we carry out an evaluation of the  melting temperature of the S phase at
$P \sigma^2/\epsilon = 0.15$ [Fig.~\ref{fig7}(c)] and $P\sigma^2/\epsilon = 7.00$ as a check on the accuracy of the coexistence line.

\subsection{L-HDT coexistence}

\begin{figure}[htp]
\centering\includegraphics[clip=true, trim=0 0 0 0, height=6.0cm, width=8.0cm]{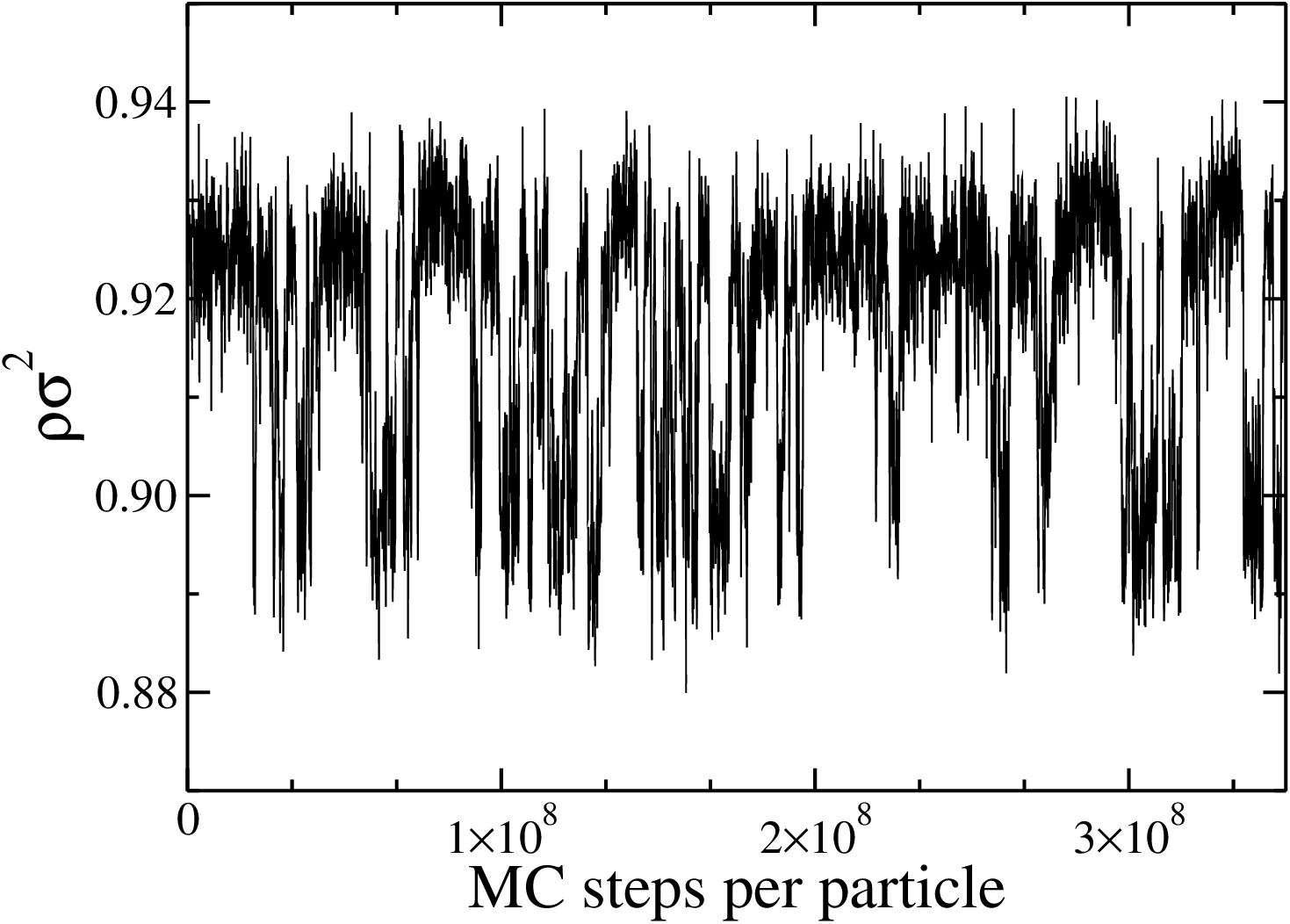}
\caption{Sample time series of the number density near the  L-HDT coexistence curve,   
with $N=986$, $k_B T/\epsilon = 5.0$ and $P_0=50.0\epsilon/\sigma^2$.
The system samples both the
(lower density) liquid and the HDT crystal.}
\label{fig8}
\end{figure}

\begin{figure}[ht]
\centering\includegraphics[clip=true, trim=0 0 0 0, height=6.0cm, width=8.0cm]{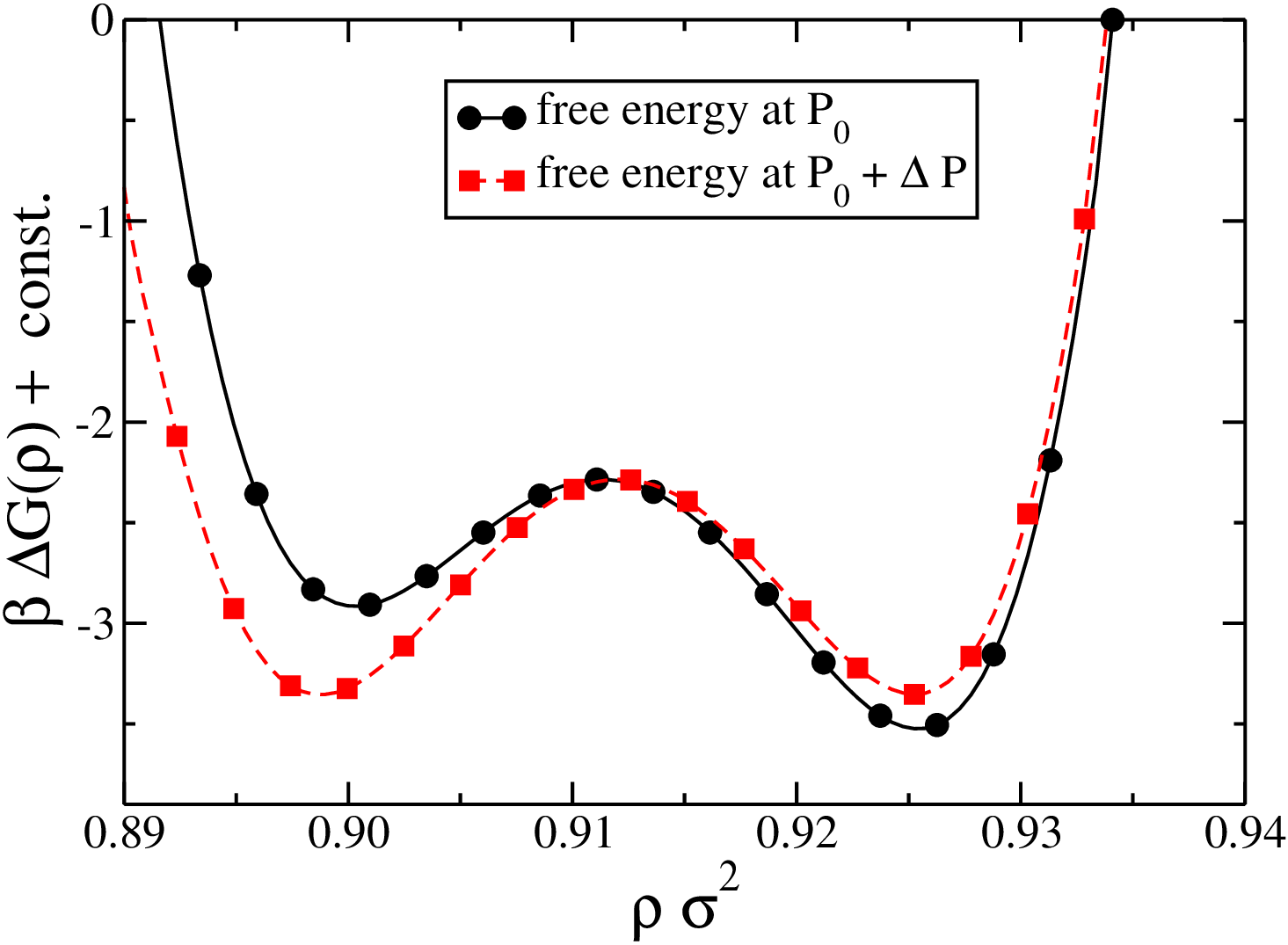}
\caption{Conditional Gibbs free energy as a function of $\rho$. At $k_B T/\epsilon=5.0$ and a pressure 
$P_0=50.0\epsilon/\sigma^2$ slightly above coexistence (solid curve, circles), 
the high density basin (HDT crystal) has a lower free energy than the low density (liquid) basin.
Through Eq.~\ref{pressure-shift}, an appropriate shift in the  pressure locates the coexistence pressure, i.e., transforms the $P_0$ curve so 
that the liquid and HDT minima are at the same level to within precision of the data (dashed line, squares).
}
\label{fig9}
\end{figure}

Using the Gibbs-Duhem integration method is not necessary (or possible) for tracing out the L-HDT melting line at high $T$, as over a certain range in $P$ the 
system can fairly easily sample both states.  Thus, to determine the coexistence $P$ along an isotherm, we first locate a pressure $P_0$
for which we can sample both states with reasonable statistics, as shown in Fig.~\ref{fig8}, and determine the conditional 
Gibbs free energy from a histogram of the densities sampled during an NPT simulation,
\begin{equation}
 \beta\Delta G(T,P_0;\rho) = -\ln{[P_r(\rho)]},
\label{free-energy}
\end{equation}
where $P_r(\rho)$ is the probability density of observing the system at a particular $\rho$.  Here, we do not normalize our histograms as the 
normalization merely adds an inconsequential shift. $P_0$ already provides an estimate of the location of the coexistence pressure.  
The conditional free energy shown in Fig.~\ref{fig9} (black curve) exhibits a global minimum at high density (HDT) and a metastable one at 
low density (liquid).  The free energy barrier between the two states is characteristic of a first order transition.
To more precisely locate the coexistence pressure, we reweight  the histogram by applying a pressure shift,
\begin{equation}
\beta \Delta G(T,P^\prime;\rho) = \beta \Delta G(T,P_0;\rho) + \frac{N \beta \Delta P}{\rho} + c,
\label{pressure-shift}
\end{equation}
where $c$ is a constant related to normalization and $\Delta P$ is the pressure shift that brings
the two minima to the same level, as in Fig.~\ref{fig9} (red curve).  The coexistence pressure is then equal to $P^\prime=P_0 + \Delta P$.
In practice, the shift we obtain is hardly perceptible on the scale of our plots, e.g., for the $k_B T/\epsilon=5.0$ case in Fig.~\ref{fig9},
$P_0 \sigma^2/\epsilon=50.0$ and $\Delta P \sigma^2/\epsilon=-0.135$, 
and for $k_B T/\epsilon=1.0$, $P_0 \sigma^2/\epsilon=14.350$ and $\Delta P \sigma^2/\epsilon=-0.004$. 
We note that the barrier does grow with decreasing $T$, and below $k_B T/\epsilon\approx 0.5$, both phases can stably exist for sufficiently
long times in order to perform Gibbs-Duhem integration.  Indeed below this $T$, it is not feasible to continue with histogram reweighting
without using some biasing potential within the MC simulations.

\subsection{G-L coexistence}

The G-L coexistence line can be determined by using the Gibbs ensemble MC method developed by Panagiotopoulos~\cite{panagiotopoulos}. 
The Gibbs ensemble employs two separated subsystems (without the presence of an interface), where the total number of particles 
is fixed and the total volume (in this case, area) of the two subsystems is also fixed; 
the total system as a whole evolves according to  the canonical ensemble. The thermodynamic requirements for phase coexistence are that the temperature, pressure, and chemical potential of the two coexisting phases must be equal and these requirements can be achieved by performing three different kinds of trial MC moves. First, particle displacement within each subsystem, second, volume fluctuations of the two subsystems, and third, transferring particles between the two subsystems. The advantage of using the Gibbs ensemble is that the system finds the densities of the coexistence phases without computing either the pressure or the chemical potential. 

Having obtained the coexistence densities at a series of $T$, the corresponding coexistence pressures can be estimated by 
applying the virtual volume change method of Haresmiadis {\it et al}~\cite{harismiadis}. In this method, we perform 
separate NVT MC simulations of both the liquid and the gas at their respective  coexistence densities (at a given $T$), and obtain the pressure via,
\begin{equation}
 P = \frac{k_B T}{\Delta V} \ln\left[\left\langle \left( \frac{V^\prime}{V} \right)^2 \exp{(-\beta \Delta U)}\right\rangle\right],
\end{equation}
where $\Delta U$ is the potential energy difference between a configuration 
with particle coordinates isotropically rescaled to accommodate a smaller 
virtual area $V^\prime$ and the unaltered configuration with original area $V$, where $V^\prime = V - \Delta V$ and $\Delta V = 0.1\sigma^2$.  
Both phases give the same pressure to within error.

However, as the temperature approaches the critical temperature $T_C$, G-L coexistence can no longer be discerned in the Gibbs ensemble simulation. 
Our data for the G-L coexistence curve from the Gibbs ensemble extend only to $k_B T/\epsilon=0.50$.  Beyond this $T$, we extrapolate
according to the following procedure.
We estimate $T_C$  by fitting the density difference of the two coexisting phases to a scaling law~\cite{frenkel1,massimo,vega3}, 
\begin{equation}
 \rho_l -\rho_g = A |T-T_C|^{\beta_c},
\label{critical_temperature}
\end{equation}
where $\beta_c$ is the critical exponent, which is equal to $0.125$ for a two-dimensional system, and $A$ is a constant determined from the fit. To estimate the critical density $\rho_{C}$, we fit our results to the law of rectilinear diameters~\cite{frenkel1,massimo,vega3},
\begin{equation}
 \frac{\rho_{l} + \rho_g}{2} = \rho_{C} + B |T-T_C|,
\label{critical_density}
\end{equation}
where $B$ is a constant determined in the fit, and $T_C$ is used from the fit in Eq.~\ref{critical_temperature}. The critical pressure $P_C$ is estimated by fitting the vapor pressure curve to the Clausius-Clapeyron equation~\cite{vega3},
\begin{equation}
 \ln P = C + \frac{D}{T},
\label{critical_pressure}
\end{equation}
where $C$ and $D$ are constants determined in the fit. $P_C$ is then calculated by substituting $T_C$ obtained from Eq.~\ref{critical_temperature} in Eq.~\ref{critical_pressure}.
From the fits, we obtain $\rho_{C}\sigma^2=0.263\pm0.002$, $k_B T_C/\epsilon=0.533\pm0.002$ and $P_C\sigma^2/\epsilon=0.019\pm0.001$.
The uncertainties quoted here are based on uncertainties in the fit parameters and do not reflect any systematic error associated with the
fact that we are extrapolating above $k_B T/\epsilon = 0.50$, the highest $T$ at which we have reliable Gibbs ensemble data.

\section{Results}

\begin{figure}
\centering
\subfigure{
\includegraphics[clip=true, trim=0 0 0 0, height=6.0cm, width=8.0cm]{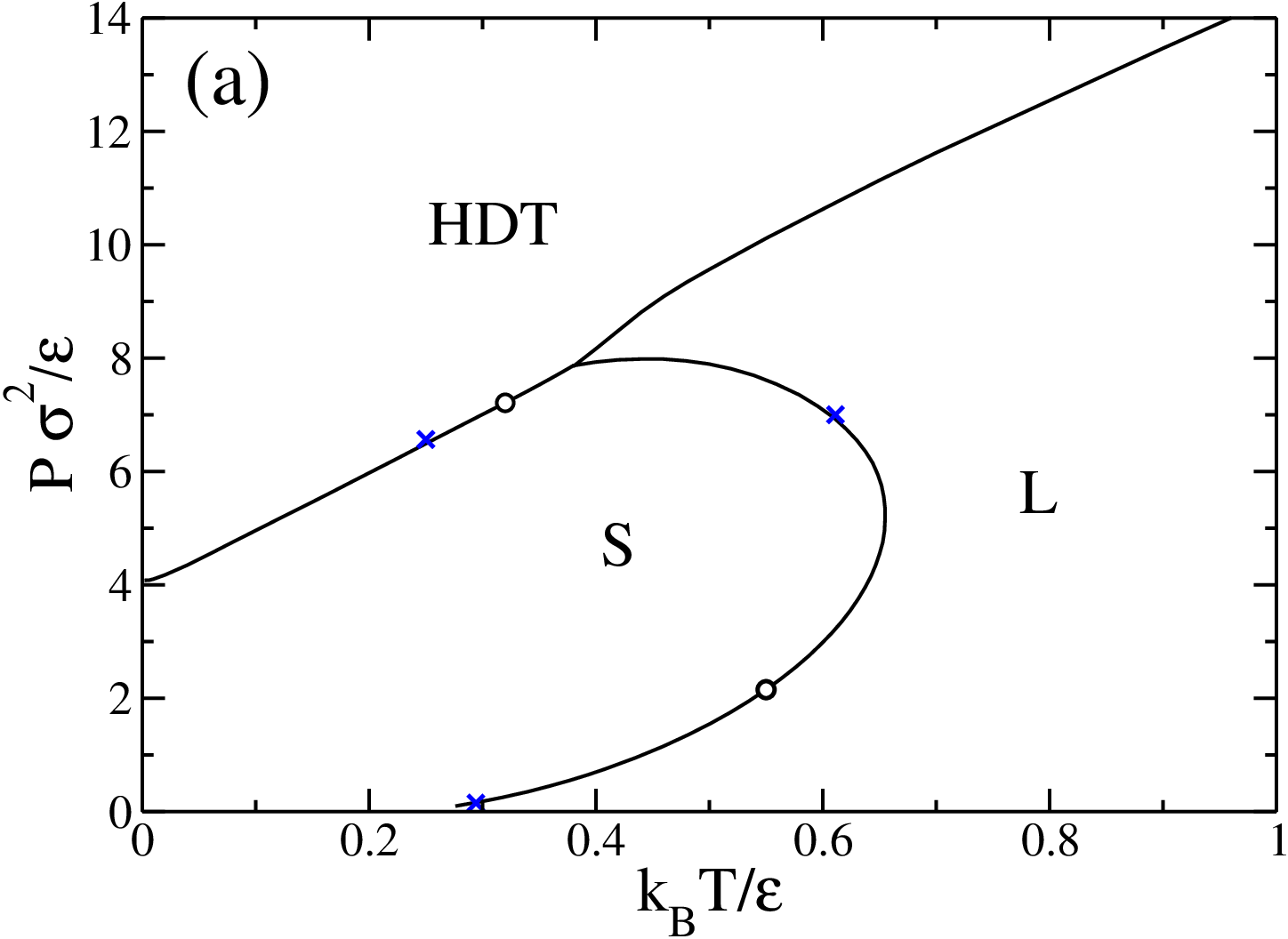}
\label{fig10a}
}
\subfigure{
\includegraphics[clip=true, trim=0 0 0 0, height=6.0cm, width=8.0cm]{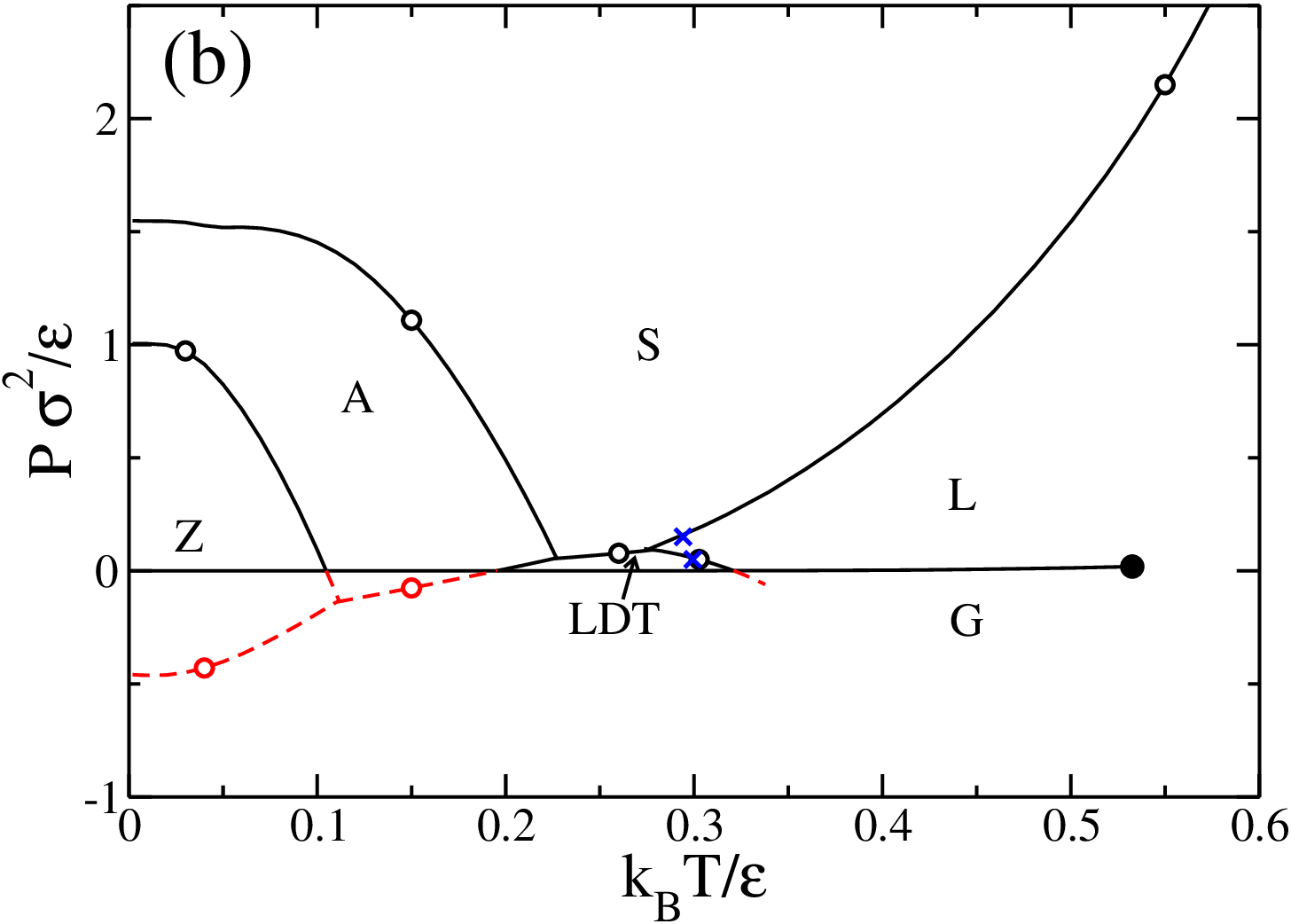}
\label{fig10b}
}
\subfigure{
\includegraphics[clip=true, trim=0 0 0 0, height=6.0cm, width=8.5cm]{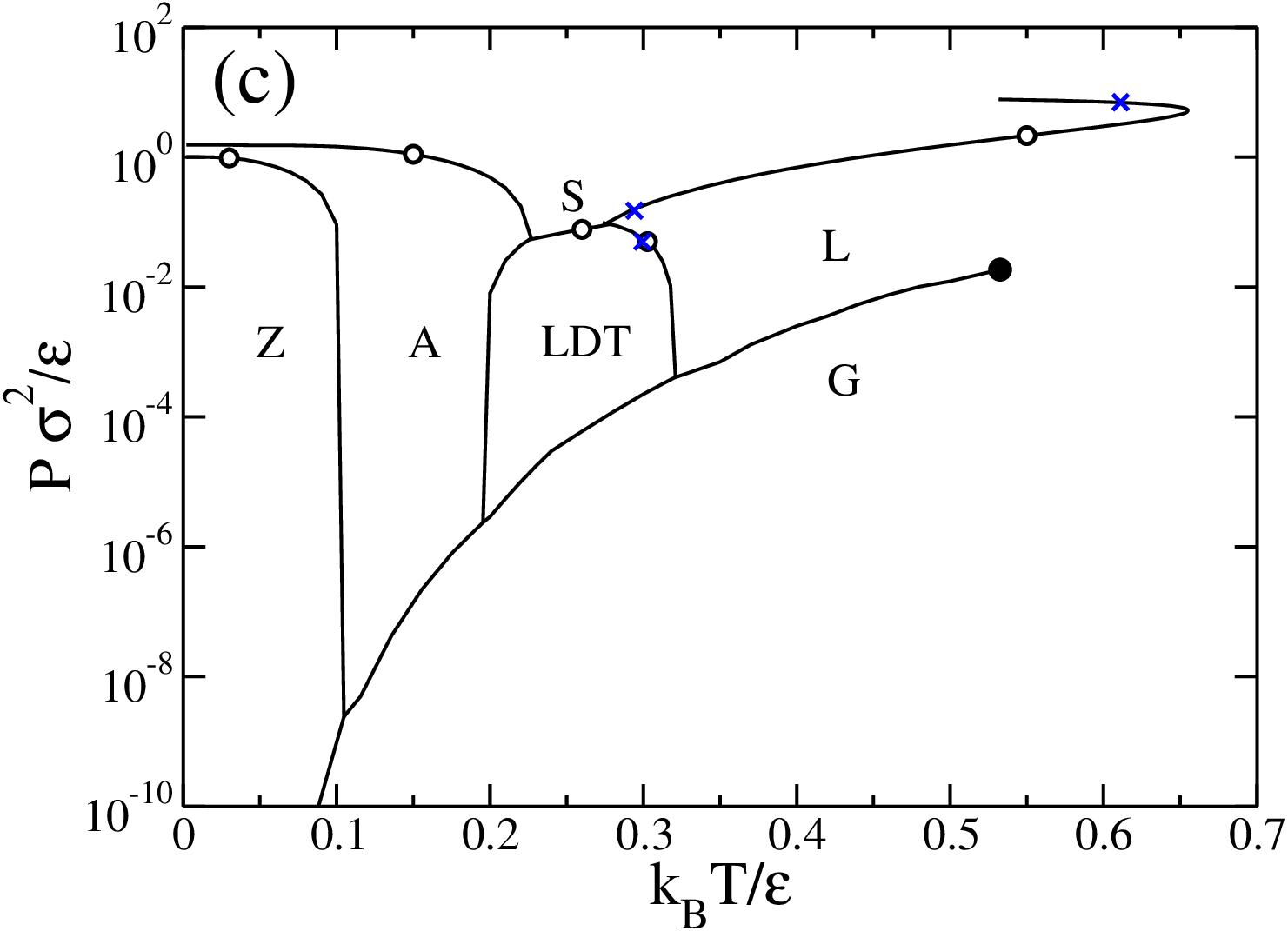}
\label{fig10c}
}
\caption{
Phase diagram of the 2D model in the $P$-$T$ plane, showing the liquid (L), gas (G) 
and crystal phases HDT, S, LDT, A and Z (see Fig.~\ref{configurations}).  The panels show portions of the phase diagram at
\subref{fig10a} high, \subref{fig10b} medium and \subref{fig10c} low $P$.
The liquid-gas coexistence line terminates at a critical point at $k_B T_c/\epsilon=0.533$ and $P_C\sigma^2/\epsilon=0.0185$ (filled circle).
Dashed lines in (b) are metastable coexistence lines assuming the absence of the gas phase. 
Initial coexistence points, i.e., starting points for Gibbs-Duhem integration, are indicated by circles, while $\times$'s show repeated
coexistence calculations done as checks on the Gibbs-Duhem integration.}
\label{ptpd}
\end{figure}

\begin{figure}[!htb]
\centering
\subfigure{
\includegraphics[clip=true, trim=0 0 0 0, height=6.0cm, width=8.0cm]{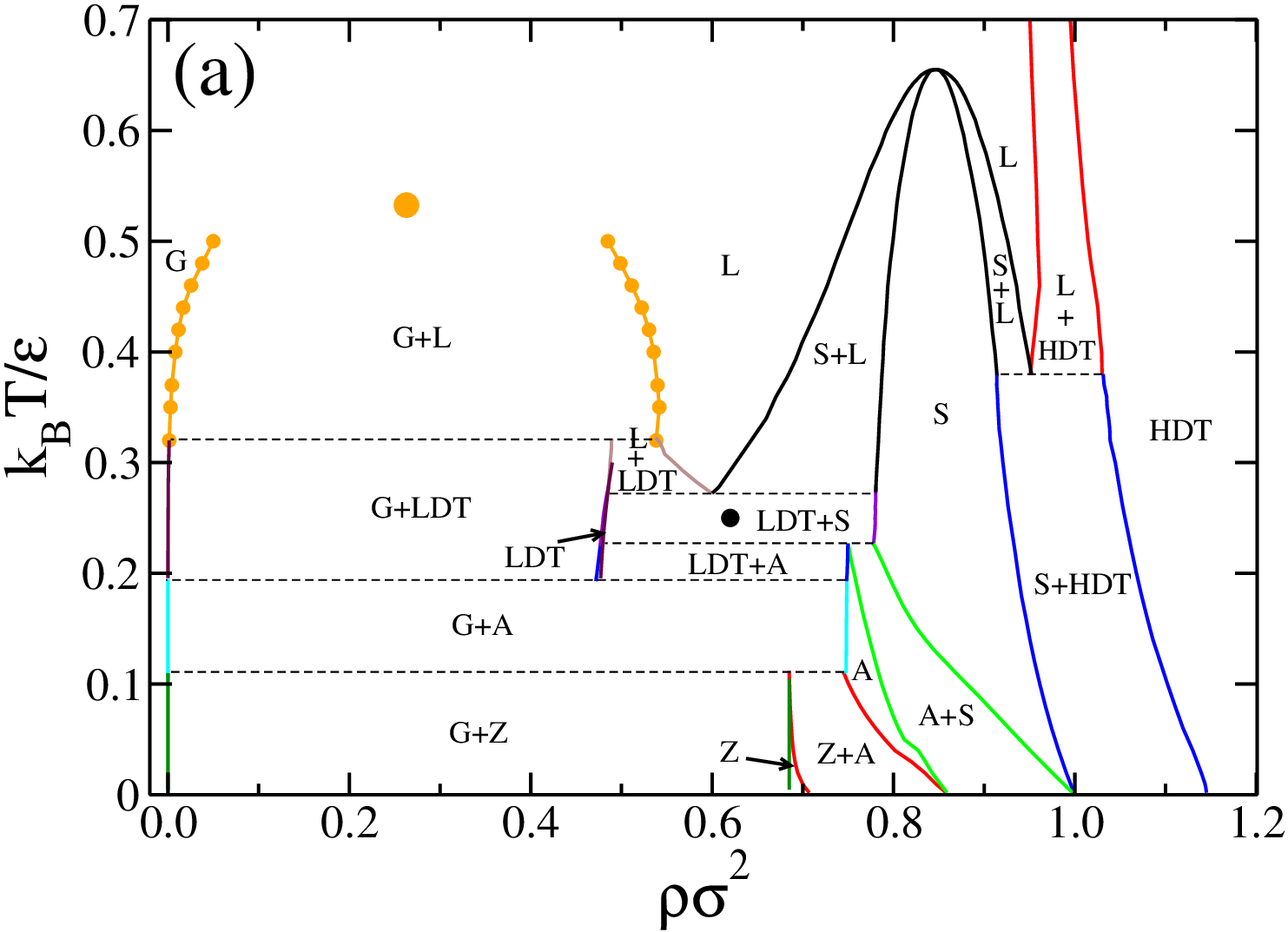}
\label{fig11a}
}
\subfigure{
\includegraphics[clip=true, trim=0 0 0 0, height=6.0cm, width=8.0cm]{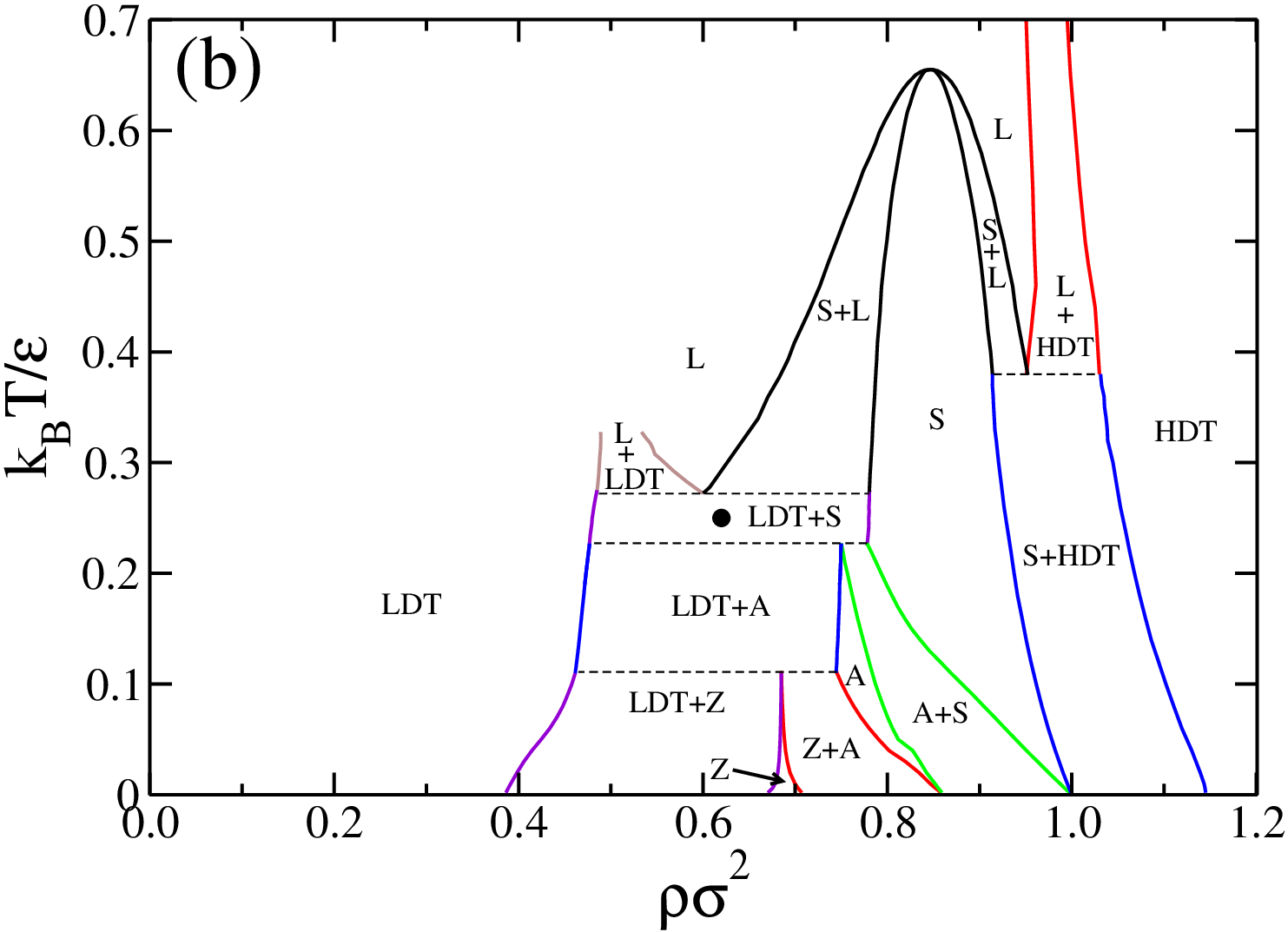}
\label{fig11b}
}
\caption{Phase diagram in the $\rho$-$T$ plane. 
(a) The points along the G-L coexistence lines indicate results from Gibbs Ensemble simulations and
the large filled orange circle shows our estimate of the G-L critical point based on an extrapolation described in the text.
Panel (b) shows the phase diagram in the absence of the gas phase.  
The filled black circle shows the location of the obscured L-L critical point discussed in Ref.~\cite{sergey}.}
\label{vtpd}
\end{figure}

Having assembled all of the individual coexistence curves, we present the  phase diagram in the $P$-$T$ plane in 
Fig.~\ref{ptpd} and in the $\rho$-$T$ plane in Fig.~\ref{vtpd}. The three panels of Fig.~\ref{ptpd} show progressively
smaller ranges of $P$.  In Fig.~\ref{fig10b}, dashed lines indicate metastable extensions of  coexistence lines
into the gas stability field (i.e., showing the phase diagram in the absence of the gas).
As an aid to interpreting Fig.~\ref{vtpd}, we recall that 
under conditions of constant volume, the thermodynamic ground state is not necessarily a single phase, but is generally  
composed of two  coexisting phases.
The bottom panel of Fig.~\ref{vtpd} shows the phase diagram in the absence of the gas phase.

\begin{figure}[htp]
\centering\includegraphics[clip=true, trim=0 0 0 0, height=6.0cm, width=8.0cm]{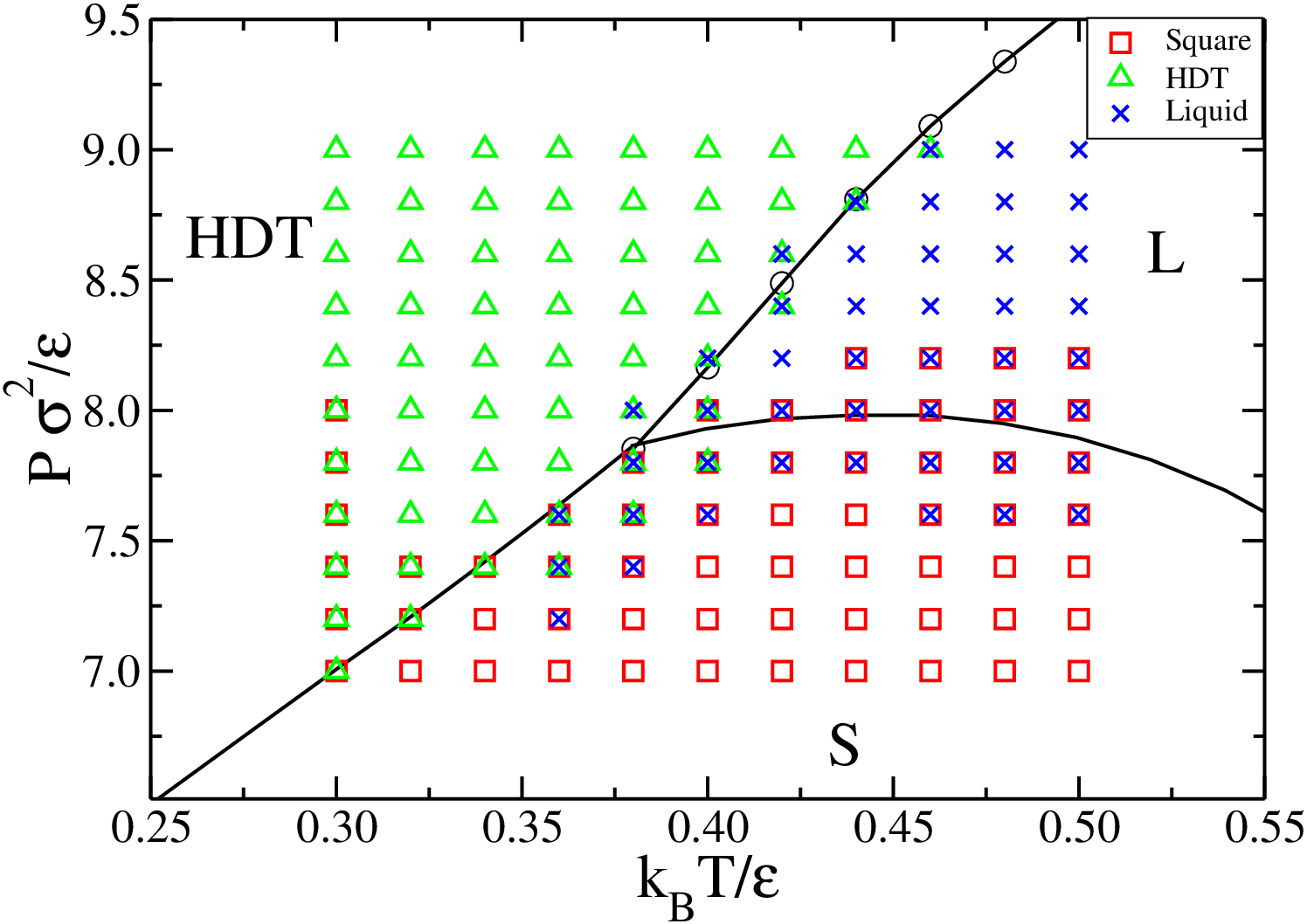}
\caption{P-T phase diagram obtained at high pressure, near the L-S-HDT triple point.
The grid of points is obtained from three sets of simulations.  Each of L, S and HDT is used to initialize a simulation set
with $N=986$, 992 and 986, respectively.
The final phase adopted from each set at each state point is indicated by a symbol: S, square; HDT, triangle; L, $\times$.
E.g., at low $P$ and high $T$, both L and HDT transform to S, while near the triple point, each phase retains metastability.}
\label{fig12}
\end{figure}

Fig.~\ref{fig10a} shows a prominent S-L melting line temperature maximum at 
$P \sigma^2/\epsilon=5.24\pm 0.05$ and $k_B T_{\rm max}/\epsilon=0.655\pm0.005$. At this point, according to Eq.~\ref{gibbsduhem}, 
the molar volumes of the S crystal and liquid are equal. At higher $P$, the melt is more dense than the crystal, as in 
the familiar case of water and hexagonal ice.

An even more exotic feature of the S-L melting line is the pressure maximum occurring near the HDT-S-L triple point
at $P_{\rm max} \sigma^2/\epsilon=7.98\pm0.08$ and $k_B T/\epsilon=0.450\pm0.003$.
A close-up of this feature is shown in Fig.~\ref{fig12}.  At this point, 
according to Eq.~\ref{gibbsduhem},
the entropy of the S crystal and the liquid are equal, and for lower $T$ along the curve, the melt has a {\it lower} entropy 
than the crystal.
The presence of the pressure maximum in the melting curve allows for ``inverse melting''~\cite{stillinger} in the narrow 
range of $P$ between the triple point and the maximum, i.e. isobaric heating of the liquid results in crystallization.

Given the numerical uncertainties in determining coexistence conditions and tracing out coexistence lines, we carry out
a rough check by preforming three sets of simulations in the vicinity of the HDT-S-L triple point.  
Each set is a grid of 121 simulations for state points marked in Fig.~\ref{fig12}. 
For one set, the particles are initially positioned on the S lattice; for the second set, points are initially on the HDT lattice;
high $T$ liquid state configurations seed the third set of simulations.  We run each simulation for $5\times 10^7$ MC steps per particle,
and then indicate with the appropriate symbol in Fig.~\ref{fig12} the phase which the
system spontaneously adopts.
Potentially, since there are three simulations per state point, three symbols may appear, indicating stability or metastability of all three phases.
Near the triple point, the simulations retain the starting phase, as expected, while
deep within a stability field, all sets transform to the same phase.
In this way, we crudely map out the extent of metastability.  

It is difficult to directly confirm inverse melting
on typical simulation time scales, as the metastable phase
is never far from the coexistence line.  We aim to address this in future work.
However,  and while this is not a definitive check on the existence of inverse
melting, the tendency for points exhibiting liquid metastability within the S stability field to track the curvature of 
the S-L melting line is supportive of the existence of this phenomenon in the system, i.e., the lowest $P$ point for each $T$ for
which the $\times$ and $\circ$ simultaneously occur roughly form a curve with a maximum in $P$ that tracks the 
shape of the S-L coexistence line.

At lower $P$, we confirm the negative slope of the LDT-L melting line as reported already in Ref.~\cite{sergey}.  Below the
LDT-S-L triple point, we find that the new crystal phases A and Z both have reasonably large stability fields, as shown in 
Fig.~\ref{fig10b}, and that the LDT
crystal, having the lowest density of the crystal phases studied, occupies a rather small portion of the phase diagram.
The A-S transition line is also negatively sloped, which together with the fact that the A phase has a lower density than S
(see Fig.~\ref{vtpd}), implies through Eq.~\ref{gibbsduhem} that the entropy of S is larger than that of A.  Indeed, the bonding
distances required to form A are rather restrictive compared to the geometry of S, and this is reflected in the smaller range in 
$\rho$ for which A is the single stable phase (again, compared to S).  A similar argument holds when comparing Z to A.

\begin{figure}[htp]
\centering\includegraphics[clip=true, trim=0 0 0 0, height=6.0cm, width=8.0cm]{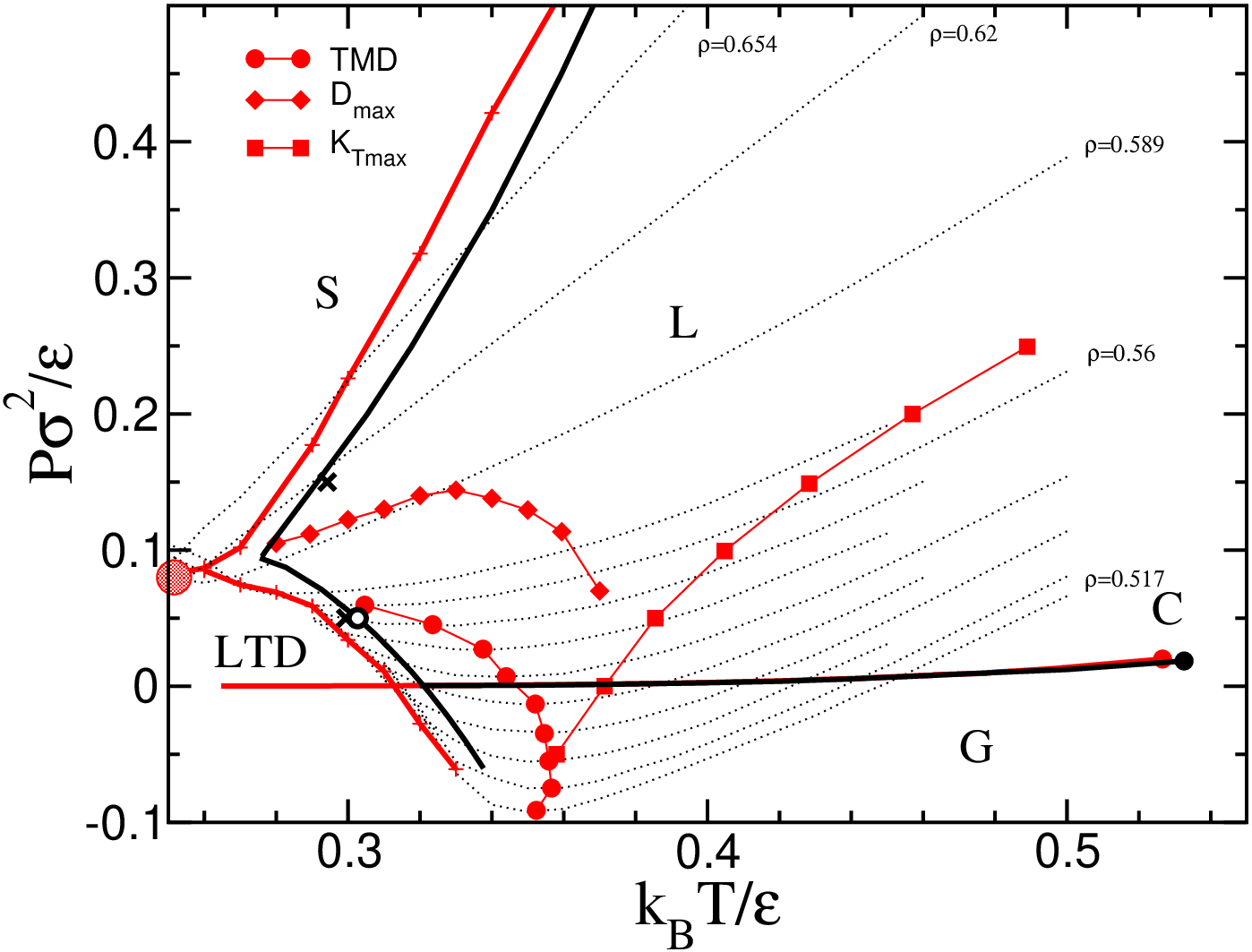}
\caption{
Comparison of our phase diagram with previously reported system properties.
Red curves are taken from Ref.~\cite{sergey} and represent crystallization lines ($+$),
locus of temperatures of maximum density along isobars (TMD, circles), pressures of 
maximum diffusivity along isotherms ($D_{\rm max}$, diamonds), maxima of isothermal compressibility
($K_{\rm Tmax}$, squares) and G-L coexistence.  Also shown are the G-L critical point (filled red circle) and
the obscured L-L critical point (large hashed circle).  
Dotted lines show pressure along  isochores.  All other symbols as in Fig.~\ref{ptpd}.
We note that we determine the location of the G-L critical point from an extrapolation of data above 
$k_B T/\epsilon=0.50$, while the one reported in Ref.~\cite{sergey} is based on inflection points of
pressure isotherms.  The previously reported crystallization lines 
fall within the presently calculated crystal stability fields.}
\label{fig13}
\end{figure}

In Ref.~\cite{sergey}, the authors locate lines in the $P$-$T$ plane that demarcate a limit to observing the liquid,
i.e., where crystallization is practically unavoidable.  Although their investigation into this aspect of the model
was not exhaustive, the character of crystallization was possibly suggestive of continuous crystallization seen in other
two dimensional systems.  We plot these lines within the appropriate portion of our calculated phase diagram in 
Fig.~\ref{fig13}.  We see that the crystallization lines occur well below our calculated first-order melting lines,
and therefore occur at conditions for which there is a gap in crystal and liquid chemical potential.  However, this does
merit a closer look at the crystallization process, especially near the apparent limit of liquid metastability.  Also in
Fig.~\ref{fig13}, we plot the location of what might be termed the obscured L-L critical point at low $T$
that appears to be responsible for the liquid anomalies reported in Ref.~\cite{sergey}, but which is unobservable
owing to unavoidable nucleation.  Within uncertainty, this obscured critical point falls on the S-LDT coexistence line.

We estimate the density of the obscured critical point from the pressure isochores reported in Ref.~\cite{sergey},
and plot the location in the $\rho$-$T$ plane in Fig.~\ref{vtpd}.  We see that it falls within the coexistence region
of two crystals of significantly different $\rho$, S and LDT.  This is similar to the case of, e.g., the TIP4P2005 model of
water~\cite{vega4,vega5} and is consistent with the idea that L-L phase separation is possible when there is a 
strong coupling between energy and density~\cite{pccp}.

\section{Discussion}

We calculate the coexistence temperature (along an isobar) or pressure (along an isotherm) of two 
phases by determining the point at which the chemical potential of those two phases cross, and estimate the 
uncertainty by accounting for the numerical error, typically arising from an integration, in evaluating the various
terms in, e.g., Eq.~\ref{muintegral}, \ref{transformed_integration} and \ref{free_energy}.  The errors mostly 
result in a constant shift in the curves that, given the small difference in slopes of $\mu(P)$ or $\mu(T)$ between  
the liquid and crystal  phases, can lead to a large uncertainty in the crossing.
As a check, after calculating the 
coexistence curve through Gibbs-Duhem integration, for the L-S case for example, we determine
two additional  chemical potential crossings along different thermodynamic paths and the results show good consistency 
with the Gibbs-Duhem curve.  
Another indicator of the quality of the results are the degree to which coexistence
lines cross at the L-S-HDT and L-S-LDT triple points.

Having said this, shifts in the $\mu( P)$ or $\mu(T)$ curves do not affect the slopes, which show in general the
first-order character of the L-S or L-LDT transitions.
For a given phase, we determine $\mu$ to the point where it is simple to determine the equilibrium properties of that phase,
i.e., to the point where spontaneous transformation does not readily occur on the timescale of simulation.
We note that for the liquid to crystal transitions, the chemical potential difference between the liquid and crystal
at which metastability is no longer easily attainable is rather small in comparison to other studies~\cite{flavio}.
Perhaps this is a feature of two-dimensional systems, but nonetheless implies a very small surface tension
if the classical description of nucleation is valid.

The L-LDT and L-S crystallization lines in Ref.~\cite{sergey}, as noted earlier, were dynamically determined as
maximal extents of the liquid's ability to exist,
and we show here that they indeed occur in the metastable liquid. 
The loss of liquid metastability prevents observing any low and high density liquids that would exist below
the proposed L-L critical point  because these limit lines radiate from the critical point towards higher $T$.
We would like to explore the process of crystallization in this vicinity.  If indeed the L-L critical point
proposed for this system is obscured by nucleation induced through critical fluctuations, studying nucleation in the present model
may  help better understanding what may be occurring in water~\cite{molinero}.

Notably, for the model at higher $P$, we provide evidence for inverse melting, 
arising from a maximum in $P$ in the L-S coexistence line.
This phenomenon is rare, and seeing evidence for it in such a simple system will allow for deeper
exploration into the basic physics surrounding it.

The freezing of the liquid to the close-packed solid, i.e., the L-HDT transition, appears to be first-order for all $T$ 
that we have explored.  For our system size, the free energy barrier between the L and HDT basins with 
$\rho$ as the order parameter at  $k_B T/\epsilon = 5.0 $ and $P=50.0\epsilon/\sigma^2$ is just above $1 k_B T$.  
In the high $T$ limit when 
the system should behave as hard disks, Mak~\cite{mak} has provided evidence that the transition should be also
first-order.  Lowering $T$, the barrier grows and reaches a value of  $\sim 2.4  k_B T$ at a simulation conducted 
on our coexistence line at
$k_B T/\epsilon = 0.50$ and $P\sigma^2/\epsilon= 9.5649$ with $N=986$,
thus becoming more strongly first-order.  An investigation into the region of HDT close to melting 
would be warranted, as Mak has shown that for hard-disks, freezing occurs to a crystal in which
an orientational order parameters scales 
with system size in a way consistent with the hexatic phase.

\section{Conclusions}

We compute a phase diagram using various free energy techniques of a two-dimensional SSSW model
that has been previously shown to exhibit liquid-state anomalies often associated with the presence of a 
metastable L-L critical point~\cite{sadr3}.  We find two low-$T$ crystal phases not previously reported. 
All transitions, including melting lines, appear to be first-order for our system size of 
$\sim1000$ particles. 
Thus, it appears that the liquid anomalies present in the system do not arise as a result of quasi-continuous 
freezing, as has been previously suggested~\cite{wilding}.
Previously reported crystallization lines fall within respective phase stability regions
reported here.
Interestingly, the difference in chemical potential between liquid and crystal phases at
 the limit where the metastable phase can be readily observed is rather small, $\beta \Delta \mu \sim 0.01$.

The L-S coexistence curve exhibits both a maximum temperature, indicating that at higher pressure the
crystal is less dense than the melt, and a pressure maximum, which means that inverse melting should occur in a
specific pressure range.  Given the scarcity of systems exhibiting inverse melting, the present model presents
the opportunity to study this rare phenomenon in more detail.

\section*{Acknowledgments}

AMA and IS-V thank NSERC for funding, ACEnet for funding and computational support and CFI for funding of
computing infrastructure. SVB acknowledges the partial support of
this research through the Dr. Bernard W. Gamson Computational Science Center at Yeshiva College.


\begin{thebibliography}{999}
\bibitem{predcorr} The algorithm used is $P_{i+1} = P_i + \Delta T (f_i + f_{i+1})/2$, where $f_i$ is the right hand side of Eq.~\ref{gibbsduhem}, the evaluation
of which requires one $NPT$ MC simulation for each phase at temperature $T_i$ and pressure $P_i$.  
Alternatively, we iterate $T_{i+1} = T_i + \Delta P (1/f_i + 1/f_{i+1})/2$.  This is especially useful in the vicinity of 
the melting temperature maximum of the S-L coexistence curve.
\bibitem{identifysolid} The solid particles have been identified using the method developed by 
Frenkel and co-workers~\cite{frenkel3}. For each particle we define the quantity $q_{lm}(i)=1/N_b(i)\sum_{j=1}^{N_{b}(i)}Y_{lm}(\hat{r}_{ij})$,
where the sum is carried out over all neighboring particles $N_b(i)$ that are within a distance $\sqrt{2}/a$, 
and $Y_{lm}(\hat{r}_{ij})$ are the spherical harmonics calculated for the normalized direction vector 
$\hat{r}_{ij}$ between the neighbors. The unit vector $\hat{r}_{ij}$ is determined by the polar 
and azimuthal angles $\theta_{ij}\equiv \pi/2$ and $\phi_{ij}$.
Since our crystallites possess square symmetry, we use $l=4$. For each pair of neighboring particles 
$i$ and $j$, we calculate the correlation $\sum_{m=-4}^4 \hat{q}_{4m}(i)\hat{q}^*_{4m}(j)$, where 
$\hat{q}_{4m}=q_{4m}(i)/[\sum_{m=-4}^{4} |q_{4m}(i)|^2]^{1/2}$ and $q^∗$ is the complex conjugate of $q$. 
If the correlation between two neighboring particles is greater than $0.9$, 
then particles $i$ and $j$ are considered to be connected. If a particle has at least 
three connected neighbors, then it is considered a solid particle.
\end{thebibliography}
\end{document}